\documentclass[journal]{IEEEtran}
\usepackage[cmex10]{amsmath}
\usepackage{amsmath}
\usepackage{amssymb}
\usepackage{algorithm}
\usepackage{algorithmic}
\usepackage{bm}
\usepackage{cite}
\usepackage[multiple]{footmisc}
\usepackage{psfrag}

\ifCLASSINFOpdf
  \usepackage[pdftex]{graphicx}
  \graphicspath{{../pdf/}{../jpeg/}}
  \DeclareGraphicsExtensions{.pdf,.jpeg,.png}
\else
\usepackage[dvips]{graphicx}
\graphicspath{{./}}
\DeclareGraphicsExtensions{.eps}
\fi
\usepackage{psfrag}

\usepackage[caption=false,font=footnotesize]{subfig}
\usepackage{color}
\usepackage{url}

\def\Herm{\operatorname{H}}
\def\Tran{\operatorname{T}}
\def\vx{\mathbf{x}}
\def\vy{\mathbf{y}}
\def\vz{\mathbf{z}}
\def\va{\mathbf{a}}
\def\hva{\mathbf{\hat a}}
\def\mA{\mathbf{A}}
\def\mI{\mathbf{I}}
\def\mJ{\mathbf{J}}

\def\vw{\mathbf{w}}
\def\vh{\mathbf{h}}
\def\hmC{\mathbf{\hat C}}
\def\vr{\mathbf{r}}
\def\vs{\mathbf{s}}
\def\hvs{\mathbf{\hat s}}

\def\hvw{\mathbf{\hat{w}}}

\def\veta{\bm{\eta}}
\def\vxi{\bm{\xi}}

\def\vtheta{\bm{\theta}}
\def\hvtheta{\bm{\hat{\theta}}}
\def\vbeta{\bm{\beta}}
\def\hvbeta{\bm{\hat\beta}}

\begin{document}
\title{Variational Bayesian Inference of Line Spectra}
\author{Mihai-Alin Badiu, Thomas Lundgaard Hansen, and Bernard Henri Fleury
\thanks{This work was supported by the research
project VIRTUOSO (funded by Intel Mobile Communications, Anite, Telenor,
Aalborg University, and the Danish National Advanced Technology Foundation) and the Danish Council for Independent Research under grant IDs DFF--5054-00212 and DFF--4005-00549.}%
\thanks{The authors are with the Department of Electronic Systems, Aalborg University, Denmark (e-mail: \{mib,tlh,bfl\}@es.aau.dk).}
}
\maketitle


\begin{abstract}
In this paper, we address the fundamental problem of line spectral estimation in a Bayesian framework.
We target model order and parameter estimation via variational inference in a probabilistic model in which the frequencies are continuous-valued, i.e., not restricted to a grid; and the coefficients are governed by a Bernoulli-Gaussian prior model turning model order selection into binary sequence detection.
Unlike earlier works which retain only point estimates of the frequencies, we undertake a more complete Bayesian treatment by estimating the posterior probability density functions (pdfs) of the frequencies and computing expectations over them.
Thus, we additionally capture and operate with the uncertainty of the frequency estimates.
Aiming to maximize the model evidence, variational optimization provides analytic approximations of the posterior pdfs and also gives estimates of the additional parameters.
We propose an accurate representation of the pdfs of the frequencies by mixtures of von Mises pdfs, which yields closed-form expectations.
We define the algorithm VALSE in which the estimates of the pdfs and parameters are iteratively updated.
VALSE is a gridless, convergent method, does not require parameter tuning, can easily include prior knowledge about the frequencies and provides approximate posterior pdfs based on which the uncertainty in line spectral estimation can be quantified.
Simulation results show that accounting for the uncertainty of frequency estimates, rather than computing just point estimates, significantly improves the performance. The performance of VALSE is superior to that of state-of-the-art methods and closely approaches the Cram\'{e}r-Rao bound computed for the true model order.
\end{abstract}


\section{Introduction}

The problem of line spectral estimation (LSE)~\cite{StoicaMoses2005}, i.e. extracting the parameters of a superposition of complex exponential functions from noisy measurements is fundamental in numerous disciplines in engineering, physics, and natural sciences. To quote a few examples, solutions to this problem have applications to range and direction estimation in sonar and radar, channel estimation in wireless communications, speech analysis, spectroscopy, molecular dynamics, power electronics, geophysical exploration.

In LSE, the original signal $\vx = (x_0,\ldots,x_{N-1})^{\operatorname{T}}\in\mathbb{C}^N$ is a superposition of $K$ complex sinusoids, i.e.,
\begin{equation}\label{eq:OrigSig}
    x_n = \sum_{k=1}^{K}  \alpha_k  e^{j\omega_k n}, 
\end{equation}
where $\alpha_k\in\mathbb{C}$ and $\omega_k\in[-\pi,\pi)$ are the complex amplitude and (angular) frequency, respectively, of the $k$th component. We are given the vector $\vy$ containing $M\leq N$ noisy measurements of those components of $\vx$ with indices in $\mathcal{M}\subseteq\{0,\ldots,N-1\}$, $|\mathcal{M}|=M$. Defining the function $\va:[-\pi,\pi)\rightarrow\mathbb{C}^M$, $\omega\to\va(\omega) = (e^{j\omega m} \mid m\in\mathcal{M})^{\operatorname{T}}$ and the vector $\bm{\epsilon}$ representing additive noise, we write
\begin{equation}\label{eq:SigModel}
    \vy = \sum_{k=1}^{K} \alpha_k \va(\omega_k) + \bm{\epsilon}.
\end{equation}
The problem of LSE involves estimating the number $K$ of sinusoidal components, also referred to as model order selection, and their associated parameters $\alpha_k$ and $\omega_k$. Even if the model order $K$ is given, LSE is still nontrivial because of the nonlinear dependency of~\eqref{eq:SigModel} on the frequencies.\footnote{When $K$ and the frequencies are given, the complex amplitudes can be easily estimated with the linear least-squares method.}

\subsection{Prior Work}
Under the assumption of known $K$, the $\omega_k$'s are traditionally estimated using the maximum-likelihood (ML) technique or subspace methods, such as~\cite{Schmidt1986,RoyKailath1989}. The ML method involves the hard task of maximizing a nonconvex function that has a multimodal shape with a sharp global maximum. The maximizer is typically searched using iterative algorithms (e.g.,~\cite{ZiskindWax1988,Feder1988,Fleury1999}) which, however, require accurate initialization and, at best, are guaranteed to converge to a local optimum. Nonetheless, the performance of the ML technique is superior to that of subspace methods, the difference being evident especially when the sample size $M$ or alternatively the signal-to-noise ratio (SNR) are small. Since $K$ is typically unknown in practice, the model order is conventionally selected based on an information criterion, which comprises a data term representing the fitting error and a penalty term that increases with the model order (see~\cite{Stoica2004} and references therein). Assuming a range of potential model orders, the parameters corresponding to each possible order are estimated using, e.g., one of the aforementioned methods. Finally, the tradeoff between fitting error and model complexity is made by selecting the configuration that minimizes the criterion. Scanning a range of model orders can be computationally expensive. Also, in non-asymptotic regimes (particularly limited $M$ or SNR), information criteria tend to provide a wrong model order. A comprehensive review of classical approaches can be found in~\cite{StoicaMoses2005}.

A more recent approach to LSE is dictionary-based model estimation, see~\cite{Austin2013} and the references therein. In this approach, nonlinear estimation of the frequencies is avoided by discretizing the range $[-\pi,\pi)$ into a finite set (grid) of samples that represent the candidate frequency estimates. The signal model~\eqref{eq:SigModel} is then approximated with a linear system comprising a so-called dictionary matrix (whose columns are given by $\va(\cdot)$ evaluated at the grid samples) and a vector of weights. Thus, the original nonlinear problem is replaced by a linear inverse problem to which a sparse solution is sought. The nonzero entries of the sparse estimate of the weight vector encode the model order and parameter estimates. There is a plethora of techniques that can be used for sparse signal representation, see the detailed survey~\cite{Tropp2010}. However, restricting the candidate frequency estimates to a discrete grid induces spectral leakage due to the model mismatch. Consequently, $\vx$ can admit only an approximately sparse representation (or may be even incompressible) in a finite dictionary~\cite{Chi2011,Duarte2013}. On the one hand, a denser grid provides a better sparse approximation and higher accuracy of frequency estimation. On the other hand, increasing the grid density makes the dictionary columns highly coherent, which might affect the sparse reconstruction capability, and boosts the computational complexity. To alleviate the mismatch issues, several approaches are conceived, e.g.: in~\cite{Duarte2013}, the concept of structured sparsity is utilized to inhibit closely-spaced frequency estimates; the method in~\cite{Malioutov2005} starts with a coarse grid and heuristically iterates between estimating the weights and placing a finer grid around the location of the non-zero weight estimates; in~\cite{Ekanadham2011,Yang2013,Hu2013,Fyhn2015}, a less fine grid is used as a baseline and the dictionary matrix is modified to include auxiliary interpolation functions.

In the quest for gridless methods which work directly with continuously parameterized dictionaries, i.e., dictionaries whose parameter ranges in $[-\pi,\pi)$, several works depart from using a static dictionary given by a fixed grid. By including the parameters that dictate the dictionary in the estimation problem, they obtain dynamic dictionary algorithms in which the candidate frequencies and hence the dictionary columns are gradually refined. In~\cite{Austin2013}, two such algorithms are designed based on the $\ell_p$ regularized least squares objective by adding a penalty term to prohibit closely spaced frequencies and respectively imposing a hard constraint on the minimum distance between frequencies. The algorithms approximately solve the involved nonlinear estimation and still require an initial grid~\cite{Austin2013}. A different line of works adopts the Bayesian framework and augments the probabilistic model of sparse Bayesian learning (SBL)~\cite{Tipping2001,Wipf2004} to incorporate the candidate frequencies. In SBL, a sparse weight vector is promoted by selecting a parameterized/hierarchical prior model for its entries~\cite{Tipping2001,Wipf2004}. Estimation in the augmented model is performed using variational inference methods~\cite{ShutinFleury2011,Hu2012,Shutin2013} or maximization of the marginalized posterior pdf~\cite{Hansen2014}. Common to all existing SBL-based approaches is that they restrict to compute point estimates of the frequencies (i.e., MAP/ML estimates), which implies nontrivial maximization of highly multimodal functions (similar to classical ML frequency estimation) in each iteration. The maximization is accomplished approximately by using a grid followed by refinement with Newton's method or interpolation. Another limitation is that, while providing good reconstruction performance, the SBL-based methods reportedly overestimate the model order, i.e., they consistently output additional spurious components (artifacts) of small power~\cite{ShutinFleury2011,Shutin2013}.

A different gridless approach that avoids the frequency discretization issues is based on the atomic norm (equivalently, the total variation norm), which is the continuous analog of the $\ell_1$ norm and allows for working with an infinite, continuous dictionary. In this way, it is shown that for the noiseless case the frequencies can be perfectly recovered from complete data ($M=N$)~\cite{Candes2014} or incomplete data ($M<N$)~\cite{Tang2013}, as long as they are well separated. In~\cite{Bhaskar2013}, the atomic norm soft thresholding (AST) method, which solves a convex program, is proposed for LSE from noisy, complete data. AST is generalized to the incomplete data case in~\cite{Yang2015}. Given that AST requires knowledge of the noise variance, the grid-based SPICE method~\cite{Stoica2011} (which minimizes a covariance matrix fitting criterion) is extended in~\cite{Yang2015} to gridless SPICE (GLS). GLS is applicable to both complete and incomplete data cases without knowledge of the noise power and is equivalent to atomic-norm-based methods; however, it has the limitations of frequency splitting and inaccurate model order estimation~\cite{Yang2015}. To overcome the two drawbacks, a GLS-based framework is proposed in~\cite{Yang2015}, in which: GLS is used as a method to estimate the covariance matrix of $\vy$, based on which the model order is selected using the SORTE algorithm~\cite{He2010} and the frequencies are estimated with MUSIC~\cite{Schmidt1986}.

An important limitation of atomic-norm-based techniques is that they require the frequencies be sufficiently separated in order to be recovered. Enhanced matrix completion (EMaC)~\cite{Chen2014} and reweighted atomic-norm minimization (RAM)~\cite{Yang2016} are two recent algorithms that are reported to improve the resolution capability of atomic-norm methods.

\subsection{Contribution}
In this paper, we propose a method for LSE from the measurement model~\eqref{eq:SigModel} by following the approach of sparse Bayesian inference including estimation of continuous-valued frequencies. The key development that sets our work apart from the related methods~\cite{ShutinFleury2011,Hu2012,Shutin2013,Hansen2014} is that, instead of retaining only point estimates of the frequencies, we seek a more complete Bayesian treatment by estimating pdfs of the frequencies and computing expectations over them. Our basic motivation is that, in general, a fully Bayesian approach is expected to show benefits, especially in the situations where sample sizes or SNRs are limited. The fully Bayesian approach naturally allows for representing and operating with the uncertainty of the frequency estimates, in addition to only that of the weights as considered so far. In particular, our approach involves computing expectations of $e^{jn\Theta}$, rather than just evaluating the phasor at a certain point estimate. The uncertainty impacts all other estimates and also the criterion for accepting a component in the estimated model (through the estimates involved) and therefore the model order estimate. Our results show that accounting for the uncertainty of frequency estimates with the fully Bayesian approach proves to be essential for improving model estimation performance. A second distinction from related works is that we employ a Bernoulli-Gaussian hierarchical model for the weights~\cite{Kormylo1982} instead of the typical SBL prior model~\cite{Tipping2001,Wipf2004}. By analyzing the component-acceptance criteria induced by the two models, we observe that the Bernoulli-Gaussian model is more resilient to insertion of small spurious components.

We provide our probabilistic formulation of LSE in Section II. Since exact inference in the proposed model requires computations that do not admit closed-form analytical expressions, we take the variational approach\footnote{Variational methods are deterministic inference techniques which provide analytical approximations of posterior pdfs, unlike the stochastic method of Markov chain Monte Carlo (MCMC) sampling. The convergence of MCMC methods can be prohibitively slow and difficult to diagnose. MCMC sampling has been previously used for LSE, see~\cite{Andrieu1999} and the references therein.} to: compute approximate posterior pdfs of the frequencies and weights; attempt MAP detection of the binary vector of the hierarchical model; and target ML estimation of the noise variance and parameters of the Bernoulli-Gaussian model. The variational optimization problem consists in maximizing a lower bound on the model evidence over the pdfs and parameters of interest. In Section III, we derive implicit expressions for local maximizers, which are to be updated iteratively.
To enable closed-form expectations over the approximate pdfs of the frequencies, we show in Section IV that these pdfs can be very well represented by mixtures of von Mises pdfs (see also Appendices B and C). In Section V, we propose a specific initialization and schedule of iterations that define the variational LSE (VALSE) algorithm. VALSE has several attractive features: it is fully automated (i.e., does not include parameters to be tuned, as all necessary parameters are learned from the data); it converges because each step increases the lower bound on the model evidence; it has the ability to easily incorporate prior knowledge about the frequencies (through a von Mises pdf or a mixture of such pdfs); it provides posterior distributions based on which uncertainty in LSE can be quantified. In Section VI the performance of VALSE is evaluated and compared against state-of-art methods through computer simulations. Finally, Section VII concludes the paper.

\section{Bayesian Formulation and Variational Approach}
Given the difficulty of not knowing the  model order $K$ in~\eqref{eq:SigModel}, for the design of our Bayesian estimator we propose a probabilistic model consisting of a superposition of $N$ (i.e. the dimension of the original signal $\vx$ in~\eqref{eq:OrigSig}) complex sinusoids that have random frequencies and weights. Since we want that eventually only $K$ of those components have nonzero weights, we use a sparsity-promoting prior model for the weights. Inference in the following model ideally would recover the $K$ true frequencies and corresponding nonzero weights and yield zero weights for the excessive $N-K$ components. Concretely, we assume that the measurement vector $\vy$ is a realization of a random process described by
\begin{equation}\label{eq:EstSigMod}
    \bm{Y} = \sum_{i=1}^{N} W_i \va(\Theta_i) + \bm{U}.
\end{equation}
The complex weights $\bm{W} = [W_1,\ldots,W_N]^{\operatorname{T}}$ are governed by independent Bernoulli variables $\bm{S} = [S_1,\ldots,S_N]^{\operatorname{T}}$ such that the elements of $\bm{W}\mid\bm{S}$ are independent and $(S_i,W_i)$ has a Bernoulli-Gaussian distribution. That is,
\begin{equation}\label{eq:pWcondS}
    p_{W_i\mid S_i}(w_i\mid s_i;\tau) = (1-s_i)\delta(w_i) + s_i f_{\text{CN}}(w_i;0,\tau)
\end{equation}
and $p_{S_i}(s_i) = \rho^{s_i}(1-\rho)^{(1-s_i)}$. Since $S_i=0$ implies that $W_i=0$, the probability $\rho$ controls how likely it is for the $i$th component to be ``active'' (i.e. its weight to be nonzero). In~\eqref{eq:pWcondS}, $W_i\mid S_i=1$ has a zero-mean Gaussian pdf with variance $\tau$.\footnote{While $p_{W_i\mid S_i}(w_i\mid s_i=1)$ should model some prior knowledge about the amplitudes, for the design of our estimator we select a zero-mean Gaussian pdf mainly for computational convenience (see Sec.~\ref{sec:Infer_W_S}). In fact, in the simulation experiments we generate the complex amplitudes in~\eqref{eq:OrigSig} from a distribution different than Gaussian.}
In this paper, $f_{\text{CN}}(\cdot;\bm{\mu},\bm{\Sigma})$ denotes the complex univariate/multivariate Gaussian pdf with mean $\bm{\mu}$ and covariance $\bm{\Sigma}$. The frequencies $\bm{\Theta} = [\Theta_1,\ldots,\Theta_N]^{\operatorname{T}}$ have the prior pdf $p_{\bm{\Theta}}(\vtheta)=\prod_i p_{\Theta_i}(\theta_i)$. As justified in Section~\ref{sec:approx_vonMises}, $p_{\Theta_i}$ is a von Mises pdf, or a mixture of such pdfs if one wants to model a more sophisticated, possibly multimodal distribution; the lack of prior knowledge can be represented by setting the concentration parameter of the von Mises pdf to zero. We assume that the components of the noise $\bm{U}$ are iid complex Gaussian with mean zero and variance $\nu$, which gives the likelihood
\begin{equation}\label{eq:likelihood}
    p_{\bm{Y}\mid\bm{\Theta},\bm{W}}(\vy\mid\vtheta,\vw;\nu) = f_{\text{CN}}(\vy;\sum_i w_i \va(\theta_i),\nu\mI).
\end{equation}
The model parameters are collectively denoted by $\vbeta = \{\nu,\rho,\tau\}$.

We can relate model~\eqref{eq:EstSigMod} to a sparse approximation problem in which, given the frequencies $\bm{\Theta}=\vtheta$, $\mA(\vtheta)=\left[\va(\theta_1)\ldots \va(\theta_N)\right]$ is the dictionary matrix and we need to infer the weights $\bm{W}$ from $M\leq N$ data samples. Using sparsity-promoting hierarchical models for $\bm{W}$ is a common Bayesian approach to find sparse solutions to ill-posed problems in compressed sensing. While the Bayesian treatment of LSE~\cite{ShutinFleury2011,Hu2012,Shutin2013,Hansen2014} typically uses the SBL prior model~\cite{Tipping2001,Wipf2004}, the Bernoulli-Gaussian model~\cite{Kormylo1982,Soussen2011} has not been used in the LSE context before. In the Bernoulli-Gaussian model, the binary vector $\bm{S} = [S_1,\ldots,S_N]^{\operatorname{T}}$ represents the support of the weights $\bm{W}$. Contrary to the standard sparse estimation problem, in our context the dictionary is parameterized by the frequencies that are to be inferred as well.

\begin{figure}[!t]
  \centering
  \psfrag{ft} [][][0.9]{$p_{\Theta_1}$}
  \psfrag{t}  [][][0.9]{$\Theta_1$}
  \psfrag{ft2}[][][0.9]{$p_{\Theta_i}$}
  \psfrag{t2} [][][0.9]{$\Theta_i$}
  \psfrag{ft3}[][][0.9]{$p_{\Theta_N}$}
  \psfrag{t3} [][][0.9]{$\Theta_N$}
  \psfrag{fy} [][][0.9]{$p_{\bm{Y}\mid\bm{\Theta},\bm{W}}$}
  \psfrag{w}  [][][0.9]{$W_1$}
  \psfrag{w2} [][][0.9]{$W_i$}
  \psfrag{w3} [][][0.9]{$W_N$}
  \psfrag{fw} [][][0.9]{$p_{W_1\mid S_1}$}
  \psfrag{fw2}[][][0.9]{$p_{W_i\mid S_i}$}
  \psfrag{fw3}[][][0.9]{$p_{W_N\mid S_N}$}
  \psfrag{g}  [][][0.9]{$S_1$}
  \psfrag{g2} [][][0.9]{$S_i$}
  \psfrag{g3} [][][0.9]{$S_N$}
  \psfrag{fg} [][][0.9]{$p_{S_1}$}
  \psfrag{fg2}[][][0.9]{$p_{S_i}$}
  \psfrag{fg3}[][][0.9]{$p_{S_N}$}
  \psfrag{V}  [][]{$\vdots$}
  \includegraphics[width=0.9\columnwidth]{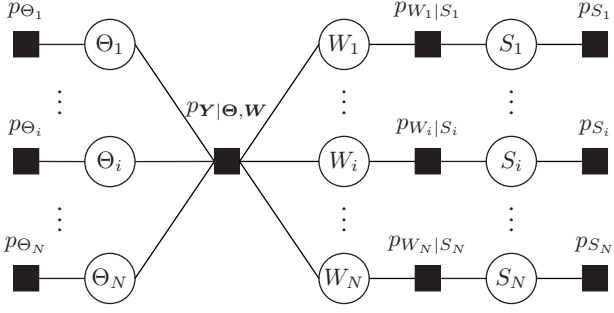}
  \caption{Factor graph representation of the joint pdf~\eqref{eq:joint_pdf}.}
  \label{fig:FacGraph}
\end{figure}

We would like to compute mean and circular mean estimates of $\bm{W}$ and $\bm{\Theta}$, respectively, based on the posterior pdf
\begin{equation}\label{eq:posterior_pdf}
    p_{\bm{\Theta},\bm{W},\bm{S}\mid\bm{Y}}(\vtheta,\vw,\vs\mid\vy;\vbeta)
    =\frac{p_{\bm{Y},\bm{\Theta},\bm{W},\bm{S}}(\vy,\vtheta,\vw,\vs;\vbeta) }
    {p_{\bm{Y}}(\vy;\vbeta)}.
\end{equation}
In~\eqref{eq:posterior_pdf}, the joint pdf in the numerator is the likelihood~\eqref{eq:likelihood} times the prior pdfs defined above, i.e.
\begin{equation}\label{eq:joint_pdf}
\begin{split}
    &p_{\bm{Y},\bm{\Theta},\bm{W},\bm{S}}(\vy,\vtheta,\vw,\vs;\vbeta) \\ &=
    p_{\bm{Y}\mid\bm{\Theta},\bm{W}}(\vy\mid\vtheta,\vw;\nu) \prod_{i=1}^N p_{\Theta_i}(\theta_i) p_{W_i\mid S_i}(w_i\mid s_i) p_{S_i}(s_i),
\end{split}
\end{equation}
while the denominator $p_{\bm{Y}}(\vy;\vbeta)$, called the model evidence (or marginal likelihood of $\vbeta$), is the marginal of the joint pdf and acts as a normalizing constant. Fig.\ref{fig:FacGraph} illustrates the factor graph representation of~\eqref{eq:joint_pdf}.
The sought estimates unfortunately require operations (high-dimensional integrals, summation over $2^N$ possible values of $\vs$) that cannot be performed analytically. Therefore we use variational inference to compute a surrogate pdf $q_{\bm{\Theta},\bm{W},\bm{S}\mid\bm{Y}}$ that should approximate~\eqref{eq:posterior_pdf} well and at the same time enable tractable estimation.

The variational approach builds on the fact that, for any postulated pdf $q_{\bm{\Theta},\bm{W},\bm{S}\mid\bm{Y}}$, the log model evidence can be expressed as~\cite[Ch. 10]{Bishop2006}
\begin{equation}\label{eq:log_evidence}
        \ln p_{\bm{Y}}(\vy;\vbeta) = \mathcal{D}_{\text{KL}}(q_{\bm{\Theta},\bm{W},\bm{S}\mid\bm{Y}} || p_{\bm{\Theta},\bm{W},\bm{S}\mid\bm{Y}}) + \mathcal{L}(q_{\bm{\Theta},\bm{W},\bm{S}\mid\bm{Y}}).
\end{equation}
The first term in~\eqref{eq:log_evidence} is the Kullback-Leibler divergence of $p_{\bm{\Theta},\bm{W},\bm{S}\mid\bm{Y}}$ from $q_{\bm{\Theta},\bm{W},\bm{S}\mid\bm{Y}}$,\footnote{The KL divergence of a pdf $p$ from a pdf $q$ (both defined on some set $\mathcal{X}$) is $\mathcal{D}_{\text{KL}}(q || p)=\int_{\mathcal{X}} q(x) \ln\frac{q(x)}{p(x)}\operatorname{d}\!x$.} while the functional $\mathcal{L}$ reads
\begin{equation}\label{eq:free_energy}
    \mathcal{L}(q_{\bm{\Theta},\bm{W},\bm{S}\mid\bm{Y}})
    =
    \operatorname{E}_{q_{\bm{\Theta},\bm{W},\bm{S}\mid\bm{Y}}} \left[\ln\frac{p_{\bm{Y},\bm{\Theta},\bm{W},\bm{S}}(\vy,\bm{\Theta},\bm{W},\bm{S};\vbeta) } {q_{\bm{\Theta},\bm{W},\bm{S}\mid\bm{Y}}(\bm{\Theta},\bm{W},\bm{S}\mid\vy)} \right].
\end{equation}
Given that $p_{\bm{Y}}(\vy;\vbeta)$ is constant w.r.t. $q_{\bm{\Theta},\bm{W},\bm{S}\mid\bm{Y}}$ and $\mathcal{D}_{\text{KL}} \geq 0$, minimizing the divergence is equivalent to maximizing $\mathcal{L}$ and tightening it as a lower bound to the log model evidence. The KL divergence vanishes only when $q_{\bm{\Theta},\bm{W},\bm{S}\mid\bm{Y}} = p_{\bm{\Theta},\bm{W},\bm{S}\mid\bm{Y}}$, in which case $\mathcal{L}$ attains its maximum value, $\ln p_{\bm{Y}}(\vy;\vbeta)$. Nonetheless, as we already mentioned, working with the posterior pdf~\eqref{eq:posterior_pdf} is intractable so we have to restrict the family of candidate pdfs.

We postulate that $q_{\bm{\Theta},\bm{W},\bm{S}\mid\bm{Y}}$ factors as
\begin{align}\label{eq:posterior_pdf_approx}
    &q_{\bm{\Theta},\bm{W},\bm{S}\mid\bm{Y}}(\vtheta,\vw,\vs\mid\vy) \nonumber \\
    &= \prod_{i=1}^N q_{\Theta_i\mid\bm{Y}}(\theta_i\mid\vy) \, q_{\bm{W}\mid\bm{Y},\bm{S}}(\vw\mid\vy,\vs)\, q_{\bm{S}\mid\bm{Y}}(\vs\mid\vy).
\end{align}
That is, we assume that the frequencies are \emph{a posteriori} independent (mutually and of the other variables).\footnote{The assumed factorization of $q_{\bm{\Theta}\mid\bm{Y}}$ is also referred to as a na\"{\i}ve mean field approximation.} Furthermore, we consider that $q_{\bm{S}\mid\bm{Y}}$ has all its mass at $\hvs$, i.e., $q_{\bm{S}\mid\bm{Y}}(\vs\mid\vy) =\delta(\vs - \hvs)$, where the function $\delta$ equals $1$ when $\mathbf{s} = \mathbf{\hat{s}}$ and $0$ otherwise. The simplifying restrictions define a family of pdfs and our goal is to search for the member which maximizes the lower bound $\mathcal{L}$.

The estimates of interest are computed from $q_{\bm{\Theta},\bm{W},\bm{S}\mid\bm{Y}}$ as follows. Since $\Theta_i$ is an angle, its estimate $\hat\theta_i$ is defined so as to give the mean direction of $e^{j\Theta_i}$~\cite{MardiaJupp2000}:
\begin{equation}\label{eq:theta_est_def}
    \hat\theta_i = \arg \left(\operatorname{E}_{q_{\Theta_i\mid\bm{Y}}} [ e^{j\Theta_i} ] \right), \quad i\in\{1,\ldots,N\}.
\end{equation}
The estimates $\operatorname{E}_{q_{\Theta_i\mid\bm{Y}}} [e^{j n\Theta_i}]$, $n\in\{0,\ldots,N-1\}$ are central in this work. Their magnitudes are $\leq 1$ with equality if, and only if $q_{\Theta_i\mid\bm{Y}}$ is the Dirac delta distribution. A broad $q_{\Theta_i\mid\bm{Y}}$ signifying high uncertainty gives a small magnitude, and vice versa. Those estimates with indices in $\mathcal{M}$ give the elements of $\hva_i=\operatorname{E}_{q_{\Theta_i\mid\bm{Y}}} [\va(\Theta_i)]$; similarly, $\|\hva_i\|_2^2\leq M$. The mean and covariance estimates of the weights are defined as
\begin{equation}\label{eq:w_est_def}
    \hvw = \operatorname{E}_{q_{\bm{W}\mid\bm{Y}}} [ \bm{W} ]
    \text{ and } \hmC=  \operatorname{E}_{q_{\bm{W}\mid\bm{Y}}} \left[ \bm{W} \bm{W}^{\Herm} \right] - \hvw\hvw^{\Herm}.
\end{equation}
Given that $q_{\bm{S}\mid\bm{Y}} = \delta(\vs - \hvs)$, the posterior pdf of $\bm{W}$ is
\begin{align}
    q_{\bm{W}\mid\bm{Y}}(\vw\mid\vy)
    &= q_{\bm{W}\mid\bm{Y},\bm{S}}(\vw\mid\vy,\hvs). \label{eq:blf_W_0}
\end{align}
Intuitively, the closer $q_{\bm{\Theta},\bm{W},\bm{S}\mid\bm{Y}}$ is to $p_{\bm{\Theta},\bm{W},\bm{S}\mid\bm{Y}}$, the better the estimates~\eqref{eq:theta_est_def} and~\eqref{eq:w_est_def} approximate the estimates which we would have computed from~\eqref{eq:posterior_pdf}, if we could. The forms of the pdfs and the support estimate $\hvs$ in the r.h.s. of~\eqref{eq:posterior_pdf_approx} result from maximizing the lower bound $\mathcal{L}$. When the parameters in $\vbeta$ are unknown, we target their ML estimates also by maximizing the lower bound to the log marginal likelihood $\ln p_{\bm{Y}}(\vy;\vbeta)$.

Finally, based on $\hvtheta$ and $\hvw$, we define the estimates of the quantities in the original superposition~\eqref{eq:OrigSig}. Let $\mathcal{S}$ be the set of indices of the non-zero components of $\vs$, i.e.
\begin{equation*}
    \mathcal{S}=\{i\mid 1\leq i\leq N, s_i=1\}.
\end{equation*}
Analogously, we define $\hat{\mathcal{S}}$ based on $\hvs$. The estimate of the model order is the cardinality of $\hat{\mathcal{S}}$:
\begin{equation}\label{eq:est_model_order}
    \hat K = |\hat{\mathcal{S}}|.
\end{equation}
We define the reconstructed signal $\mathbf{\hat x}\triangleq (\hat x_1,\ldots,\hat x_N)^{\operatorname{T}}$ as the expectation of the signal part in the r.h.s. of~\eqref{eq:EstSigMod} over $q_{\bm{\Theta},\bm{W},\bm{S}\mid\bm{Y}}$, which gives
\begin{align}\label{eq:ReconSig}
    \hat x_n = \sum_{i\in \hat{\mathcal{S}}}  \hat w_i  \operatorname{E}_{q_{\Theta_i\mid\bm{Y}}} [e^{j n\Theta_i}], \quad n\in\{1,\ldots,N\}.
\end{align}
The components of $\hvtheta$ and $\hvw$ with indices in $\hat{\mathcal{S}}$ give the estimates of the frequencies and amplitudes in~\eqref{eq:OrigSig}.

\section{Solution to the Variational Optimization Problem}\label{sec:SolVarOpt}
We now turn to maximizing the lower bound $\mathcal{L}(q_{\bm{\Theta},\bm{W},\bm{S}\mid\bm{Y}})$ in~\eqref{eq:free_energy} with $q_{\bm{\Theta},\bm{W},\bm{S}\mid\bm{Y}}$ of the form~\eqref{eq:posterior_pdf_approx}. Except for restricting $q_{\bm{S}\mid\bm{Y}}$ to give probability one to some sequence $\hvs$, we do not impose any constraints on the forms of the factors in~\eqref{eq:posterior_pdf_approx}. That is, the forms of the approximate posterior pdfs result from variational optimization and are dictated by the likelihood~\eqref{eq:likelihood} and prior pdfs. As maximizing $\mathcal{L}$ over all factors simultaneously is not viable, we perform alternating optimization: $\mathcal{L}$ is maximized over each of the factors $q_{\bm{W},\bm{S}\mid\bm{Y}}$, $q_{\Theta_i\mid\bm{Y}}$, $i=1,\ldots,N$, in turn while keeping the others fixed. Consequently, the form of each factor is implicit because it depends on the other factors.

Upon their initialization, we iteratively cycle through the factors and replace them one by one with a revised expression. Such a scheme is guaranteed to converge to some local maximum of $\mathcal{L}$~\cite[Ch. 10]{Bishop2006}. In the following we derive the factor expressions that correspond to the fixed-point of the scheme. A specific initialization and scheduling of updates are proposed in Sec.~\ref{sec:VALSE}.

\subsection{Inferring the frequencies $\bm{\Theta}$}
For each $i=1,\ldots,N$, maximizing $\mathcal{L}$ in~\eqref{eq:free_energy} w.r.t. the factor $q_{\Theta_i\mid\bm{Y}}$ gives~\cite[Ch. 10, p. 466]{Bishop2006}
\begin{align*}
    \ln q_{\Theta_i\mid\bm{Y}}(\theta_i\mid\vy) &=  \operatorname{E}_{\sim\theta_i} \left[ \ln p_{\bm{Y},\bm{\Theta},\bm{W},\bm{S}}(\vy,\theta_i,\bm{\Theta}_{\sim i},\bm{W},\bm{S};\vbeta) \right] \\
    &\phantom{=}+ \text{const.}
\end{align*}
where the expectation is taken over $q_{\bm{W},\bm{S}\mid\bm{Y}}\prod_{j\neq i} q_{\Theta_j\mid\bm{Y}}$, the joint pdf $p_{\bm{Y},\bm{\Theta},\bm{W},\bm{S}}$ is given by~\eqref{eq:joint_pdf} and the constant ensures normalization of the pdf. We further write only the terms that depend on $\theta_i$, i.e.
\begin{align*}
    \ln q_{\Theta_i\mid\bm{Y}}(\theta_i\mid\vy) &=  \operatorname{E}_{\sim\theta_i} \left[ \ln p_{\bm{Y}\mid\bm{\Theta},\bm{W}}(\vy\mid\theta_i,\bm{\Theta}_{\sim i},\bm{W};\nu) \right]  \\
    &\phantom{=}+ \ln p_{\Theta_i}(\theta_i) + \text{const.}
\end{align*}
Plugging the Gaussian form of the likelihood~\eqref{eq:likelihood} in the above expression and carrying out the required expectations, we finally obtain
\begin{equation}\label{eq:blf_theta}
    q_{\Theta_i\mid\bm{Y}}(\theta_i\mid\vy) \propto p_{\Theta_i}(\theta_i)\exp\left\{\Re\left(\veta_i^{\Herm} \va(\theta_i)\right) \right\}
\end{equation}
where the complex vector $\veta_i$ is given by
\begin{equation}\label{eq:eta_1}
    \veta_i = \frac{2}{\nu}  \left(\vy - \sum_{l\in \hat{\mathcal{S}} \setminus \{i\}} \hat{w}_l \hva_l \right)\hat{w}_i^* - \frac{2}{\nu} \sum_{l\in \hat{\mathcal{S}} \setminus \{i\}} \hat C_{l,i} \hva_l
\end{equation}
when $i\in \hat{\mathcal{S}}$, and $\veta_i = \mathbf{0}$ otherwise. The second factor in the r.h.s. of~\eqref{eq:blf_theta} is an approximation of the marginal likelihood of $\theta_i$; it is an extremely multimodal function, see Sec.~\ref{sec:approx_vonMises}. According to~\eqref{eq:eta_1}, the likelihood favors values of $\theta_i$ for which the angle between $\hat w_i\va(\theta_i)$ and the residual signal (after canceling the interference from the other components) is small.\footnote{The angle $\phi$ between two complex vectors $\mathbf{u}$ and $\mathbf{v}$ satisfies $\cos(\phi) = \frac{\Re(\mathbf{u}^{\Herm}\mathbf{v})}{\|\mathbf{u}\|\|\mathbf{v}\|}$.}
Interestingly, the likelihood corresponds to coherent estimation of $\Theta_i$ from the residual signal when the weight is fixed to $\hat{w}_i$. At the same time, it penalizes (to an extent given by the cross-variance of the weights) values that result in small angle between $\va(\theta_i)$ and $\hva_l$ of the other components in the model. Naturally, when $i\notin \hat{\mathcal{S}}$ (i.e. $\hat s_i=0$), only the prior information comes into play in~\eqref{eq:blf_theta}.

The pdf~\eqref{eq:blf_theta} does not yield analytic expressions for $\operatorname{E}_{q_{\Theta_i\mid\bm{Y}}}[\va(\Theta_i)]$. In Section~\ref{sec:approx_vonMises}, we show that $q_{\Theta_i\mid\bm{Y}}$ in ~\eqref{eq:blf_theta} is well approximated by a mixture of von Mises pdfs, which gives a closed-form approximation of $\operatorname{E}_{q_{\Theta_i\mid\bm{Y}}}[\va(\Theta_i)]$.

\subsection{Inferring the weights $\bm{W}$ and support $\bm{S}$}\label{sec:Infer_W_S}
We next maximize $\mathcal{L}$  w.r.t. $q_{\bm{W},\bm{S}\mid\bm{Y}}(\vw,\vs\mid\vy)$ when $q_{\Theta_i\mid\bm{Y}}$, $i=1,\ldots,N$, are kept fixed. Since $q_{\bm{W},\bm{S}\mid\bm{Y}}(\vw,\vs\mid\vy)$ is restricted in~\eqref{eq:posterior_pdf_approx} to give the marginal pmf $q_{\bm{S}\mid\bm{Y}}(\vs\mid\vy)=\delta(\vs-\hvs)$, we cannot anymore use the factor-update expression corresponding to free-form optimization~\cite[Ch. 10, p. 466]{Bishop2006}. So we will explicitly carry out the maximization of $\mathcal{L}$.

Plugging the postulated pdf~\eqref{eq:posterior_pdf_approx} in~\eqref{eq:free_energy} we obtain
\begin{align*}
    \mathcal{L}(q_{\bm{W}\mid\bm{Y},\bm{S}},\hvs)
    &= \text{const.} -
    \operatorname{E}_{q_{\bm{W}\mid\bm{Y},\bm{S}}}
    \Big\{
        \ln q_{\bm{W}\mid\bm{Y},\bm{S}}(\bm{W}\mid\vy,\hvs)
    \Big. \\
    &\phantom{=}-
    \Big.
        \operatorname{E}_{q_{\bm{\Theta}\mid\bm{Y}}}
        \left[
            \ln p_{\bm{Y},\bm{\Theta},\bm{W},\bm{S}}(\vy,\bm{\Theta},\bm{W},\hvs;\vbeta)
        \right]
    \Big\}.
\end{align*}
Let us introduce the pdf
\begin{equation*}
    t(\vw;\hvs) = \frac{1}{Z(\hvs)} \exp\left\{\operatorname{E}_{q_{\bm{\Theta}\mid\bm{Y}}} \left[ \ln p_{\bm{Y},\bm{\Theta},\bm{W},\bm{S}}(\vy,\bm{\Theta},\vw,\hvs;\vbeta) \right]\right\}
\end{equation*}
where $p_{\bm{Y},\bm{\Theta},\bm{W},\bm{S}}$ is given by~\eqref{eq:joint_pdf} and $Z(\hvs)$ is the normalizing constant obtained by integrating the exponential over $\vw$. We can now write
\begin{equation}\label{eq:free_en_weights}
    \mathcal{L}(q_{\bm{W}\mid\bm{Y},\bm{S}},\hvs) = -\mathcal{D}_{\text{KL}}(q_{\bm{W}\mid\bm{Y},\bm{S}} || t) + \ln Z(\hvs) + \text{const.}
\end{equation}
Inspecting~\eqref{eq:free_en_weights}, for any $\hvs$ the maximum of $\mathcal{L}$ over $q_{\bm{W}\mid\bm{Y},\bm{S}}$ is attained when the KL divergence vanishes. Thus, $\mathcal{L}$ has its maximum at
\begin{equation}\label{eq:blf_W}
    q_{\bm{W}\mid\bm{Y},\bm{S}}(\vw\mid\vy,\hvs) = t(\vw;\hvs)\quad\text{and}\quad \hvs = \operatorname*{arg\,max}_{\vs} \ln Z(\vs).
\end{equation}
To compute $\operatorname{E}_{q_{\bm{\Theta}\mid\bm{Y}}} \left[ \ln p_{\bm{Y},\bm{\Theta},\bm{W},\bm{S}}(\vy,\bm{\Theta},\vw,\vs;\vbeta) \right]$ required for $t(\vw;\hvs)$ and $Z(\vs)$ in~\eqref{eq:blf_W}, we use~\eqref{eq:joint_pdf}, together with~\eqref{eq:likelihood} and~\eqref{eq:pWcondS}, and obtain an expression that is quadratic in $\vw$, given that all $p_{W_i\mid S_i}(w_i\mid s_i=1)$ are Gaussian. We define the matrix $\mJ$ with elements $J_{ii} = M$ and $J_{ij}=\hva_i^{\Herm}\hva_j$, $i,j=1,\ldots,N$, $j\neq i$, and the vector $\vh=\left[\hva_1^{\Herm}\vy,\ldots,\hva_N^{\Herm}\vy\right]^{\Tran}$. From~\eqref{eq:blf_W_0} and~\eqref{eq:blf_W}, we obtain
\begin{equation*}
    q_{\bm{W}\mid\bm{Y}}(\vw\mid\vy) = f_{\text{CN}}\left(\vw_{\hat{\mathcal{S}}};\hvw_{\hat{\mathcal{S}}},\hmC_{\hat{\mathcal{S}}}\right) \prod_{i\notin \hat{\mathcal{S}}} \delta(w_i),
\end{equation*}
where the mean and covariance matrix of the Gaussian posterior pdf of $\bm{W}_{\hat{\mathcal{S}}}$ are
\begin{equation}\label{eq:W_mean_var}
\hvw_{\hat{\mathcal{S}}} = \nu^{-1} \hmC_{\hat{\mathcal{S}}} \vh_{\hat{\mathcal{S}}} \quad\text{and}\quad
\hmC_{\hat{\mathcal{S}}} = \nu \left(\mJ_{\hat{\mathcal{S}}}+\frac{\nu}{\tau}\mI\right)^{-1}.
\end{equation}
The mean is the LMMSE estimate of $\bm{W}_{\hat{\mathcal{S}}}$ assuming $\vs=\hvs$. For $i\notin \hat{\mathcal{S}}$, the measurements are noninformative and, conveniently, $q_{W_i\mid\bm{Y}}(w_i\mid\vy)=p_{W_i\mid S_i}(w_i\mid s_i=0)=\delta(w_i)$, i.e., $\hat{w}_i=0$.

From~\eqref{eq:blf_W}, the sequence $\hvs$ (which determines $\hat{\mathcal{S}}$) is the maximizer of
\begin{align}\label{eq:Z}
    \ln Z(\vs)
    &= \ln\det\left(\mJ_{\mathcal{S}}+\frac{\nu}{\tau}\mI\right)^{-1} + \frac{1}{\nu}\vh_{\mathcal{S}}^{\Herm} \left(\mJ_{\mathcal{S}}+\frac{\nu}{\tau}\mI\right)^{-1} \vh_{\mathcal{S}}\nonumber \\
    &\phantom{=} + \|\vs\|_0 \ln\frac{\rho\nu}{(1-\rho)\tau}  + \text{const.}
\end{align}
Since maximizing the nonlinear function~\eqref{eq:Z} is NP-hard, in Appendix~\ref{sec:approx_maxZ} we propose a suboptimal procedure which is guaranteed to converge to a local maximum of $\ln Z(\vs)$.

\begin{figure}
\centering
\includegraphics[width=0.8\linewidth]{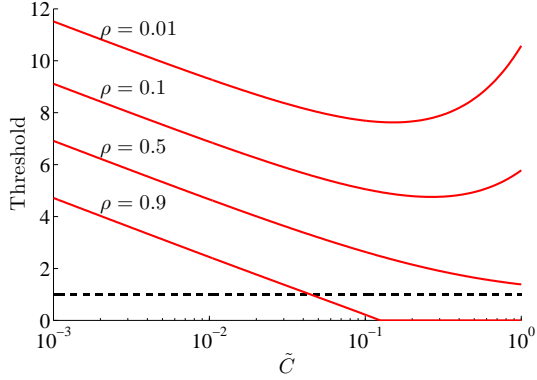}
\caption{Activation thresholds vs. weight variance for: the Bernoulli-Gaussian model (solid curves) with $\tau = 1$ and different values of $\rho$, and for the SBL prior model (dashed line). The activation test is satisfied by the points lying above the given curve.
}
\label{fig:ActivTest}
\end{figure}

According to Appendix~\ref{sec:approx_maxZ}, a sinusoidal component (we drop the index $i$ for the moment) is admitted only if the posterior mean $\hat{w}$ and variance $\hat{C}$ of its weight (for $\hat{s}=1$) satisfy
\begin{equation}\label{eq:cond_useful}
    \frac{|\hat{w}|^2}{\hat{C}} > \ln\left(\tau/\hat{C}\right) + \ln\frac{1-\rho}{\rho}.
\end{equation}
It is interesting to relate~\eqref{eq:cond_useful} to the test $|\tilde{w}|^2/\tilde{C}>1$ obtained in~\cite{Shutin2011} for the SBL prior model of the weights~\cite{Tipping2001,Wipf2004}, where $\tilde{w}$ and $\tilde{C}$ are the mean and variance of the posterior divided by the prior. The SBL prior model is often used for estimating superimposed signals~\cite{ShutinFleury2011,Hu2012,Shutin2013,Hansen2014} and, reportedly, the resulting estimators output additional spurious components (artifacts) of small power. Since $|\tilde{w}|^2/\tilde{C}$ can be viewed as an SNR of the component, the threshold can be heuristically increased such that a higher $\text{SNR}$ is required~\cite{Shutin2011,ShutinFleury2011}. For the Bernoulli-Gaussian prior model we express~\eqref{eq:cond_useful} as
\begin{equation}\label{eq:cond_useful2}
    \frac{|\tilde{w}|^2}{\tilde{C}} > \left(1+\tilde{C}/\tau\right) \ln\left[\left(1+\tau/\tilde{C}\right) \frac{1-\rho}{\rho}\right]
\end{equation}
where we used $p_{W\mid S}(w\mid s=1) = f_{\text{CN}}(w;0,\tau)$, which gives $\hat{C}^{-1}=\tilde{C}^{-1}+\tau^{-1}$ and $\hat{C}^{-1}\hat{w}=\tilde{C}^{-1}\tilde{w}$. Thus, the threshold~\eqref{eq:cond_useful2} is not constant but depends on $\rho$, $\tau$ and also $\tilde{C}$. The latter dependence makes the method more resilient to insertion of artifacts, because, as shown in Fig.~\ref{fig:ActivTest}, the threshold increases with smaller variance, unlike for the SBL model where it stays the same.

\subsection{Estimating the model parameters}\label{sec:EstParam}
The noise variance $\nu$ is often unknown in practice. Also, it might be unclear how to set the parameters $\rho$ and $\tau$ of the Bernoulli-Gaussian prior model. We show that learning the parameters can be easily included in the variational approach.

The lower bound~\eqref{eq:free_energy} now additionally depends on $\vbeta = \{\nu,\rho,\tau\}$. We alternate between maximizing $\mathcal{L}(q_{\bm{\Theta},\bm{W},\bm{S}\mid\bm{Y}},\vbeta)$ over $q_{\bm{\Theta},\bm{W},\bm{S}\mid\bm{Y}}$ for $\vbeta$ fixed to $\hvbeta$ (according to the previous subsections) and over $\vbeta$ for fixed $q_{\bm{\Theta},\bm{W},\bm{S}\mid\bm{Y}}$. In the latter step,
\begin{equation*}
    \mathcal{L}(\vbeta)
    = \operatorname{E}_{q_{\bm{\Theta},\bm{W},\bm{S}\mid\bm{Y}}} \left[\ln p_{\bm{Y},\bm{\Theta},\bm{W},\bm{S}}(\vy,\bm{\Theta},\bm{W},\bm{S};\vbeta) \right]+ \text{const.}
\end{equation*}
where we write only the term depending on $\vbeta$. The joint pdf and the approximate posterior pdf are given by~\eqref{eq:joint_pdf} and~\eqref{eq:posterior_pdf_approx}, respectively. Based on the forms of the likelihood~\eqref{eq:likelihood} and prior pdfs defined in Sec. II, we obtain
\begin{align*}
    \mathcal{L}(\vbeta)
    &= \frac{1}{\nu} \left[
    2\Re\left( \hvw_{\hat{\mathcal{S}}}^{\Herm} \vh_{\hat{\mathcal{S}}} \right) - \hvw_{\hat{\mathcal{S}}}^{\Herm} \mJ_{\hat{\mathcal{S}}} \hvw_{\hat{\mathcal{S}}} - \vy^{\Herm}\vy - \operatorname{tr}\left(\mJ_{\hat{\mathcal{S}}}\hmC_{\hat{\mathcal{S}}}\right)  \right]
    \\
    &\phantom{=}
    - M\ln\nu - \frac{1}{\tau} \left[ \hvw_{\hat{\mathcal{S}}}^{\Herm} \hvw_{\hat{\mathcal{S}}} + \operatorname{tr}\left(\hmC_{\hat{\mathcal{S}}}\right) \right] - \|\hvs\|_0\ln\tau \\
    &\phantom{=} + \|\hvs\|_0\ln\rho + (N - \|\hvs\|_0)\ln(1-\rho) + \text{const.}
\end{align*}
We can carry out $\operatorname*{arg\,max}_{\vbeta} \mathcal{L}(\vbeta)$ independently over each parameter. Equating the partial derivatives to zero gives unique solutions that correspond to the global maximum (the second-order derivatives are strictly negative). Specifically, we obtain
\begin{align}\label{eq:est_noise_var}
    \hat\nu &= \frac{1}{M} \|\vy - \sum_{i\in\hat{\mathcal{S}}} \hat w_i \hva_i\|_2^2 + \frac{1}{M} \operatorname{tr}\left(\mJ_{\hat{\mathcal{S}}}\hmC_{\hat{\mathcal{S}}}\right) \nonumber \\
    &\phantom{=} + \sum_{i\in\hat{\mathcal{S}}} |\hat w_i |^2 \left(1 -  \|\hva_i\|_2^2/M \right).
\end{align}
Thus, $\hat\nu$ takes into account not only the fitting error, but also the uncertainty of weight estimation (through $\hmC_{\hat{\mathcal{S}}}$) and of frequency estimation (via $\hva_i$). Regarding the latter, the sharper $q_{\Theta_i\mid\bm{Y}}$, the closer $\|\hva_i\|_2^2$ is to $M$ and therefore the smaller the contribution to $\hat\nu$. For $\rho$ and $\tau$ we obtain the estimates
\begin{equation}\label{eq:est_rho_tau}
    \hat\rho = \frac{\|\hvs\|_0}{N}
    \qquad\text{and}\qquad
    \hat\tau = \frac{\hvw_{\hat{\mathcal{S}}}^{\Herm} \hvw_{\hat{\mathcal{S}}} + \operatorname{tr}\left(\hmC_{\hat{\mathcal{S}}}\right) }{\|\hvs\|_0}.
\end{equation}
Naturally, $\hat\rho$ is given by the number of nonzero components of $\hvs$ and $\hat\tau$ is the averaged second-moment of the weights corresponding to those components.

\section{Approximating $q_{\Theta_i\mid\bm{Y}}$ by a mixture of von Mises pdfs}\label{sec:approx_vonMises}
In this section, after providing some background on the von Mises distribution, we show that any pdf of the form $\exp\left( \Re\left( \veta^{\Herm} \va(\theta) \right) \right)$, such as $q_{\Theta_i\mid\bm{Y}}$ in~\eqref{eq:blf_theta}, can be well represented by a mixture of von Mises pdfs (MVM). The proposed approximation enables easy computation of expectations over such pdfs. We exploit the MVM approximation in the initialization of our algorithm as well, since the exponential of the periodogram also has the said form.

\subsection{The von Mises distribution}\label{app:VM}
Among the distributions on the unit circle, the von Mises (VM) distribution is of significant importance, its role being similar to that of the Gaussian distribution on the line~\cite{MardiaJupp2000}. The pdf of the VM distribution of a random angle $\Theta$ is
\begin{equation*}
    f_\text{VM}(\theta;\mu,\kappa) = \frac{1}{2\pi I_0(\kappa)} e^{\kappa\cos(\theta-\mu)}.
\end{equation*}
The parameters $\mu$ and $\kappa$ are the mean direction and concentration parameter, respectively, and $I_p(\cdot)$ is the modified Bessel function of the first kind and order $p$. The pdf is symmetrical about its single mode, which is at $\Theta=\mu$. The VM pdf can also be parameterized in terms of $\eta = \kappa e^{j\mu}$:
\begin{equation*}
    f_\text{VM}(\theta;\eta) = \frac{1}{2\pi I_0(|\eta|)} \exp\left(\Re\{\eta^{\ast}e^{j\theta}\}\right).
\end{equation*}
The properties of circular distributions are completely determined by the characteristic function, $\varphi_p \triangleq \mathbb{E}[e^{jp\Theta}]$, $p\in\mathbb{Z}$~\cite{MardiaJupp2000}.
The characteristic function of the VM distribution is
\begin{equation}\label{eq:charfunc_VM}
    \varphi_p = e^{jp\mu}\frac{I_p(\kappa)}{I_0(\kappa)}, \quad p\in\mathbb{Z}.
\end{equation}
The moments of circular distributions are the moments of $e^{j\Theta}$, i.e., values of the characteristic function. The first moment of the VM distribution, $\varphi_1 = e^{j\mu} A(\kappa)$, gives the mean direction $\mu$ and the mean resultant length $A(\kappa) = I_1(\kappa)/I_0(\kappa)$.

The multiplication of two VM pdfs gives
\begin{align}\label{eq:prodVM}
    f_\text{VM}(\theta;\eta_1) f_\text{VM}(\theta;\eta_2)
    &\propto f_\text{VM}(\theta;\eta)
\end{align}
with $\eta=\eta_1+\eta_2$. That is, the result is proportional to a VM pdf with mean direction $\arg (\eta_1+\eta_2)$ and concentration $|\eta_1+\eta_2|$. Thus, the family of VM pdfs is closed under multiplication.

\subsection{The proposed MVM approximation}
In the following, we drop the frequency index $i$ for convenience. We write~\eqref{eq:blf_theta} as
\begin{equation}\label{eq:blf_theta_2}
    q_{\Theta\mid\bm{Y}}(\theta\mid\vy) \propto p_{\Theta}(\theta) \prod_{m\in\mathcal{M}} \exp\left(\Re\{\eta_{m}^{\ast}e^{j m \theta}\}\right),
\end{equation}
where the entries of $\veta$ have the polar form $\eta_{m}=\kappa_{m}\, e^{j\mu_{m}}$. When $0\in\mathcal{M}$ the factor in~\eqref{eq:blf_theta_2} corresponding to $m=0$ is a constant, so we can just remove this index from $\mathcal{M}$. Also, when $1\in\mathcal{M}$, the factor indexed by $m=1$ has the form of a von Mises (VM) pdf $f_\text{VM}(\theta; \eta_{1})$ with mean direction $\mu_{1}$ and concentration $\kappa_{1}$. Furthermore, the factors indexed by $m> 1$ have the form of $m$-fold wrapped VM pdfs. Thus, we can write~\eqref{eq:blf_theta_2} as
\begin{equation}\label{eq:blf_theta_3}
    q_{\Theta\mid\bm{Y}}(\theta\mid\vy) \propto p_{\Theta}(\theta)\prod_{m\in\mathcal{M}} f_\text{VM}(m \theta; \eta_{m}).
\end{equation}

\begin{figure}
\centering
\includegraphics[width=0.8\linewidth]{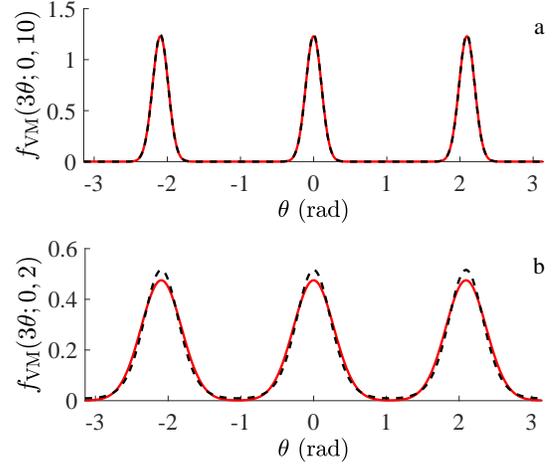}
\caption{Illustration of the approximation~\eqref{eq:approx_wrapVM}. The pdf $f_\text{VM}(3\theta;\kappa\,e^{j\cdot0})$ of a 3-fold wrapped VM distribution (dashed curve) is approximated by a mixture of $3$ von Mises pdfs (solid curve); in (a) $\kappa=10$ for which~\eqref{eq:transeq} gives $\tilde\kappa\approx 85.78$; in (b) $\kappa=2$ for which~\eqref{eq:transeq} gives $\tilde\kappa\approx 13.02$. }
\label{fig:Approx_WMV_MVM}
\end{figure}

In Appendix~\ref{app:approx} we show that a wrapped VM pdf can be very well approximated by an appropriate MVM obtained by matching their characteristic functions. Employing the result~\eqref{eq:approx_pdfs_WMV_MVM}, we approximate each of the $m$-fold wrapped VM pdfs in~\eqref{eq:blf_theta_3} by a mixture of $m$ VM pdfs, i.e.,
\begin{equation}\label{eq:approx_wrapVM}
    f_\text{VM}(m \theta;\eta_{m}) \simeq \sum_{r=0}^{m-1} \frac{1}{m} f_\text{VM}(\theta;\tilde\eta_{m,r}),
\end{equation}
where $\tilde\eta_{m,r} = \tilde\kappa_{m} e^{j\tilde\mu_{m,r}}$. The $m$ components of the MVM 
have equal amplitudes and concentrations. The value $\tilde\kappa_{m}$ of the latter is the solution to
\begin{equation}\label{eq:transeq}
    \frac{I_{m}(\tilde\kappa_{m})}{I_0(\tilde\kappa_{m})} = \frac{I_1(\kappa_{m})}{I_0(\kappa_{m})}
\end{equation}
where $I_p(\cdot)$ is the modified Bessel function of the first kind and order $p$. We show in Appendix~\ref{app:approx} that an approximate solution to the transcendental equation~\eqref{eq:transeq} can be easily found. The means $\tilde\mu_{m,r}$, $r=0,\ldots,m-1$, are given by
\begin{equation}\label{eq:means_fact_m}
    \tilde\mu_{m,r} = \frac{\mu_{m}+2\pi r}{m},
\end{equation}
i.e., they are evenly distributed around the circle, $2\pi/m$ apart. The higher the concentration parameter of the wrapped VM pdf, the better its approximation~\eqref{eq:approx_wrapVM}. As illustrated in Fig.~\ref{fig:Approx_WMV_MVM}, the approximation is very tight even for moderate values of the concentration and still good for small concentrations.

The proposed approximation enables us to exploit the fact that the family of VM pdfs is closed under multiplication. To that end, we conveniently choose the prior pdf of $\Theta$ to be $p_{\Theta}(\theta) = f_\text{VM}(\theta; \eta_{\text{a}})$, with $\eta_{\text{a}} = \kappa_{\text{a}} e^{j\mu_{\text{a}}}$.\footnote{When we do not have any prior information about $\Theta$, we can set the concentration $\kappa_{\text{a}} = 0$, in which case the prior pdf becomes the uniform circular pdf $p_{\Theta}(\theta)=1/(2\pi)$.}\footnote{Alternatively, we can select an MVM prior, if we wish to use a multimodal distribution.}
Replacing~\eqref{eq:approx_wrapVM} in~\eqref{eq:blf_theta_3} we obtain that $q_{\Theta\mid\bm{Y}}(\theta\mid\vy)$ is an MVM. Specifically, let us write $\mathcal{M}=\{m_1,m_2,\ldots,m_M\}\subseteq \{1,\ldots,N-1\}$ and define $\mathcal{R} = \{1,\ldots,m_1\}\times\ldots\times \{1,\ldots,m_M\}$. Using the multi-index $\vr=(r_1,\ldots,r_M)\in\mathcal{R}$, we have
\begin{equation}\label{eq:blf_theta_app}
     q_{\Theta\mid\bm{Y}}(\theta\mid\vy) = \frac{1}{Z_\theta} \sum_{\vr\in\mathcal{R}} \exp\left\{\Re\left(\xi_{\vr}^{\ast} e^{j\theta} \right) \right\}
\end{equation}
with
\begin{equation}\label{eq:eta_r}
    \xi_{\vr} = \eta_{\text{a}} + \tilde\eta_{m_1,r_1} + \ldots + \tilde\eta_{m_M,r_M}
\end{equation}
and the normalizing constant $Z_\theta = 2\pi \sum_{\vr\in\mathcal{R}} I_0(|\xi_{\vr}|)$. We explicitly express~\eqref{eq:blf_theta_app} as an MVM where the amplitude, mean and concentration of each of the mixture's components are given by the corresponding parameter $\xi_{\vr}$:
\begin{equation}\label{eq:blf_theta_MvM}
    q_{\Theta\mid\bm{Y}}(\theta\mid\vy) = \sum_{\vr\in\mathcal{R}} \frac{2\pi I_0(|\xi_{\vr}|)}{Z_\theta}  f_\text{VM}(\theta; \xi_{\vr}).
\end{equation}
The number $m_1\times\ldots\times m_M$ of components in~\eqref{eq:blf_theta_MvM} can be intractable. For the component with index $\vr$ to have an important contribution to~\eqref{eq:blf_theta_MvM}, its amplitude and concentration must be high, i.e., $|\xi_{\vr}|$ be large. Based on the observation that only a small fraction of them contribute significantly to the mass of $q_{\Theta\mid\bm{Y}}$, in the following we propose two heuristic methods for representing~\eqref{eq:blf_theta_MvM} by a limited number of components.

\subsection{Heuristic~1}
The first heuristic is a greedy procedure aiming to find and represent $q_{\Theta\mid\bm{Y}}$ by only the $D$ most dominant components in~\eqref{eq:blf_theta_MvM}. The idea is to progressively construct an approximation of~\eqref{eq:blf_theta_2} by sweeping through the index set $\mathcal{M}$ and including in the approximation one additional index in each step. In step $p$, $1\leq p\leq M$, we have a ``partial'' posterior pdf given by the factors in~\eqref{eq:blf_theta_2} with indices $\{m_1,\ldots,m_p\}$, i.e., only $p$ measurements are taken into account. The partial pdf is an MVM with $m_1\times\ldots\times m_p$ components parameterized by $\eta_{\text{a}} + \tilde\eta_{m_1,r_1} + \ldots + \tilde\eta_{m_p,r_p}$. As outlined in Algorithm~\ref{alg:Heuristic1}, in each step the heuristic procedure retains from the ``partial'' posterior (at most) $D$ components having the highest concentration parameters. The complexity of the greedy search is $\mathcal{O}(DMN)$. The algorithm outputs the $D$ parameters in $\vxi$ which give
\begin{equation}\label{eq:blf_theta_MvM_D}
    q_{\Theta\mid\bm{Y}}(\theta\mid\vy) \approx \sum_{d=1}^D \frac{2\pi I_0(|\xi_d|)}{\tilde Z_\theta}  f_\text{VM}(\theta; \xi_d)
\end{equation}
where $\tilde Z_\theta = 2\pi\sum_{d=1}^D I_0(|\xi_{d}|)$. Now we can compute expectations in closed-form. Using~\eqref{eq:blf_theta_MvM_D} and~\eqref{eq:charfunc_VM}, we obtain
\begin{equation*}
    \hva = \frac{2\pi}{\tilde Z_\theta } \sum_{d=1}^D \operatorname{diag} \left( I_{m_1}(|\xi_d|),\ldots, I_{m_M}(|\xi_d|) \right) \va(\arg(\xi_d)).
\end{equation*}
Similarly, the frequency estimate $\hat\theta$ defined in~\eqref{eq:theta_est_def} is given by
\begin{equation*}
    \hat\theta = \arg\left( \frac{2\pi}{\tilde Z_\theta } \sum_{d=1}^D I_1(|\xi_d|) e^{j \arg(\xi_d)} \right).
\end{equation*}

\begin{algorithm}[!t]
\caption{Heuristic~1}
\label{alg:Heuristic1}
\begin{algorithmic}[1]
\renewcommand{\algorithmicrequire}{\textbf{Input:}}
\renewcommand{\algorithmicensure}{\textbf{Output:}}
\REQUIRE $\mathcal{M}$, $\veta$, $\eta_0$ and $D$
\ENSURE  $\vxi$
\STATE{Compute all $\tilde\eta_{m,r} = \tilde\kappa_{m} e^{j\tilde\mu_{m,r}}$ in~\eqref{eq:transeq} and~\eqref{eq:means_fact_m} }
\STATE{$\vxi^{(1)}\gets(\eta_{\text{a}}+\tilde\eta_{m_1,r}\mid 0\leq r\leq m_1-1)$}
\FOR{$p=2$ to $M$}
    \STATE{$\vxi^{(p)}\gets D$ elements of $\left\{\xi^{(p-1)}_d+\tilde\eta_{m_p,r}\right\}_{d,r}$ with largest magnitudes\footnotemark}
\ENDFOR
\RETURN $\vxi=\vxi^{(M)}$
\end{algorithmic}
\end{algorithm}
\footnotetext{$\vxi^{(p)}$ has less than $D$ components when $m_1\times\ldots\times m_p < D$.}

\subsection{Heuristic~2}
The second approach is to search for the most dominant component of the mixture~\eqref{eq:blf_theta_MvM}, i.e., that with index $\operatorname*{arg\,max}_{\vr\in\mathcal{R}} |\xi_{\vr}|$. Then, we represent~\eqref{eq:blf_theta_MvM} by a single von Mises pdf based on a second-order Taylor approximation around the mean $\bar\theta$ of that component. The intuition is that, with sufficient SNR and number $M$ of measurements, the pdf~\eqref{eq:blf_theta_MvM} would peak somewhere in the neighborhood of $\bar\theta$.

Given that for each $m$, $|\tilde\eta_{m,r}|$ does not depend on $r$, see~\eqref{eq:approx_wrapVM}, to maximize $|\xi_{\vr}|$ we have to look for that $\vr$ for which the phases of the terms of~\eqref{eq:eta_r} are best aligned. Such an alignment is searched in a greedy way by Algorithm~\ref{alg:Heuristic2} whose complexity is $\mathcal{O}(MN)$. Without loss of generality we assume $m_1>\ldots>m_M$. The algorithm maintains a number $m_1$ of candidates and proceeds in a progressive manner. In step $p$, the $l$th candidate $\xi_l^{(p)}$, $1\leq l\leq m_1$, is obtained by adding the term whose phase is closest to that of $\xi_l^{(p-1)}$, i.e., having the index
\begin{equation*}
    r_l^{(p)} = \operatorname*{arg\,max}_{0\leq r\leq m_p-1} \left|\xi_l^{(p-1)} + \tilde\kappa_{m_p} \exp\left(j \tfrac{\mu_{m_p}+2\pi r}{m_p}\right) \right|.
\end{equation*}
The closed-form update is given by lines $5$ and $6$ of Algorithm~\ref{alg:Heuristic2} where $[\cdot]$ is the nearest-integer function.
We set $\bar\theta = \arg\xi_{l^\star}^{(M)}$ with $l^\star = \operatorname*{arg\,max}_{1\leq l\leq m_1} |\xi_l^{(M)}|$. Denoting the exponent of~\eqref{eq:blf_theta_2} by $f(\theta) = \Re\left(\eta_{\text{a}}^{\ast}\, e^{j\theta}+\sum_{m\in\mathcal{M}} \eta_{m}^{\ast}\, e^{j m\theta} \right)$, we make a second-order Taylor approximation of $f(\theta)$ around $\bar\theta$. Then we use the properties of the wrapped normal distribution~\cite[p. 50]{MardiaJupp2000} and its similarity to the von Mises distribution~\cite[p. 38]{MardiaJupp2000} to arrive at
\begin{equation*}
    q_{\Theta\mid\bm{Y}}(\theta\mid\vy) \approx f_\text{VM}(\theta; \hat\eta)
\end{equation*}
with $\hat\eta = \hat\kappa e^{j\hat\theta}$, $\hat\theta = \bar\theta - \frac{f'(\bar\theta)}{f''(\bar\theta)}$ and $\hat\kappa = A^{-1}\left(e^{0.5/f''(\bar\theta)}\right)$. A useful approximation of the inverse of the function $A(\cdot)=I_1(\cdot)/I_0(\cdot)$ is given in~\cite[pp. 85--86]{MardiaJupp2000}. Finally, we can easily obtain the expected value of $\va(\Theta)$,
\begin{equation*}
    \hva = \operatorname{diag} \left( \tfrac{I_{m_1}(\hat\kappa)}{I_0(\hat\kappa)},\ldots, \tfrac{I_{m_M}(\hat\kappa)}{I_0(\hat\kappa)} \right) \va(\hat\theta) .
\end{equation*}

\begin{algorithm}[!t]
\caption{Heuristic~2}
\label{alg:Heuristic2}
\begin{algorithmic}[1]
\renewcommand{\algorithmicrequire}{\textbf{Input:}}
\renewcommand{\algorithmicensure}{\textbf{Output:}}
\REQUIRE $\mathcal{M}$, $\veta$ and $\eta_0$
\ENSURE  $\hat\theta$ and $\hat\kappa$
\STATE{Compute all $\tilde\eta_{m,r} = \tilde\kappa_{m} e^{j\tilde\mu_{m,r}}$ in~\eqref{eq:transeq} and~\eqref{eq:means_fact_m} }
\STATE{$\vxi^{(1)}\gets(\eta_{\text{a}}+\tilde\eta_{m_1,r}\mid 0\leq r\leq m_1-1)$}
\FOR{$p=2$ to $M$}
    \FOR{$l=1$ to $m_1$}
        \STATE{$r_l^{(p)} = \left[ \frac{m_p \arg\left( \xi_l^{(p-1)} \right) - \mu_{m_p}}{2\pi} \right]$}
        \STATE{$\xi_l^{(p)} \gets \xi_l^{(p-1)} + \tilde\kappa_{m_p}\exp\left( j \frac{\mu_{m_p}+2\pi r_l^{(p)}}{m_p} \right)$}
    \ENDFOR
\ENDFOR
\STATE{Determine $l^\star = \operatorname*{arg\,max}_{l} |\xi_l^{(M)}|$ and set $\bar\theta = \arg \xi_{l^\star}^{(M)}$}
\RETURN $\hat\theta = \bar\theta - \frac{f'(\bar\theta)}{f''(\bar\theta)}$ and $\hat\kappa = A^{-1}\left(\exp\left(0.5/f''(\bar\theta)\right)\right)$
\end{algorithmic}
\end{algorithm}

\subsection{Illustrative examples}
To illustrate the effectiveness of the proposed approximation, we give a few simple examples where exact pdfs of $\Theta$ of the form $\exp\left\{ \Re\left( \veta^{\Herm} \va(\theta) \right) \right\}$, i.e., as in~\eqref{eq:blf_theta}, occur.

\subsubsection{Coherent estimation}
\begin{figure*}[!t]
\centering
\subfloat[]{
    \includegraphics[width=0.32\textwidth]{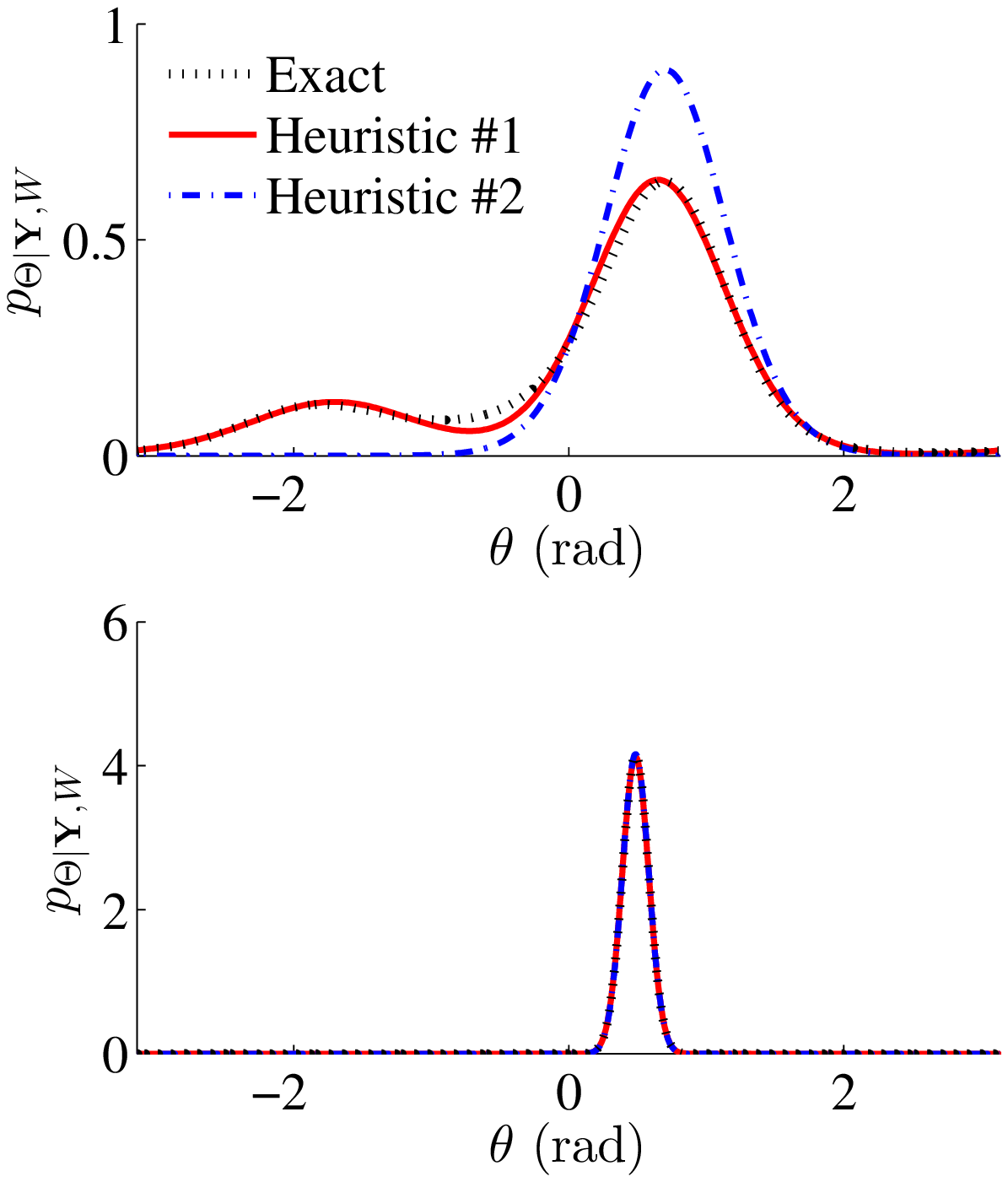}
    \label{fig:Example_coherent1}
}
\subfloat[]{
    \includegraphics[width=0.32\textwidth]{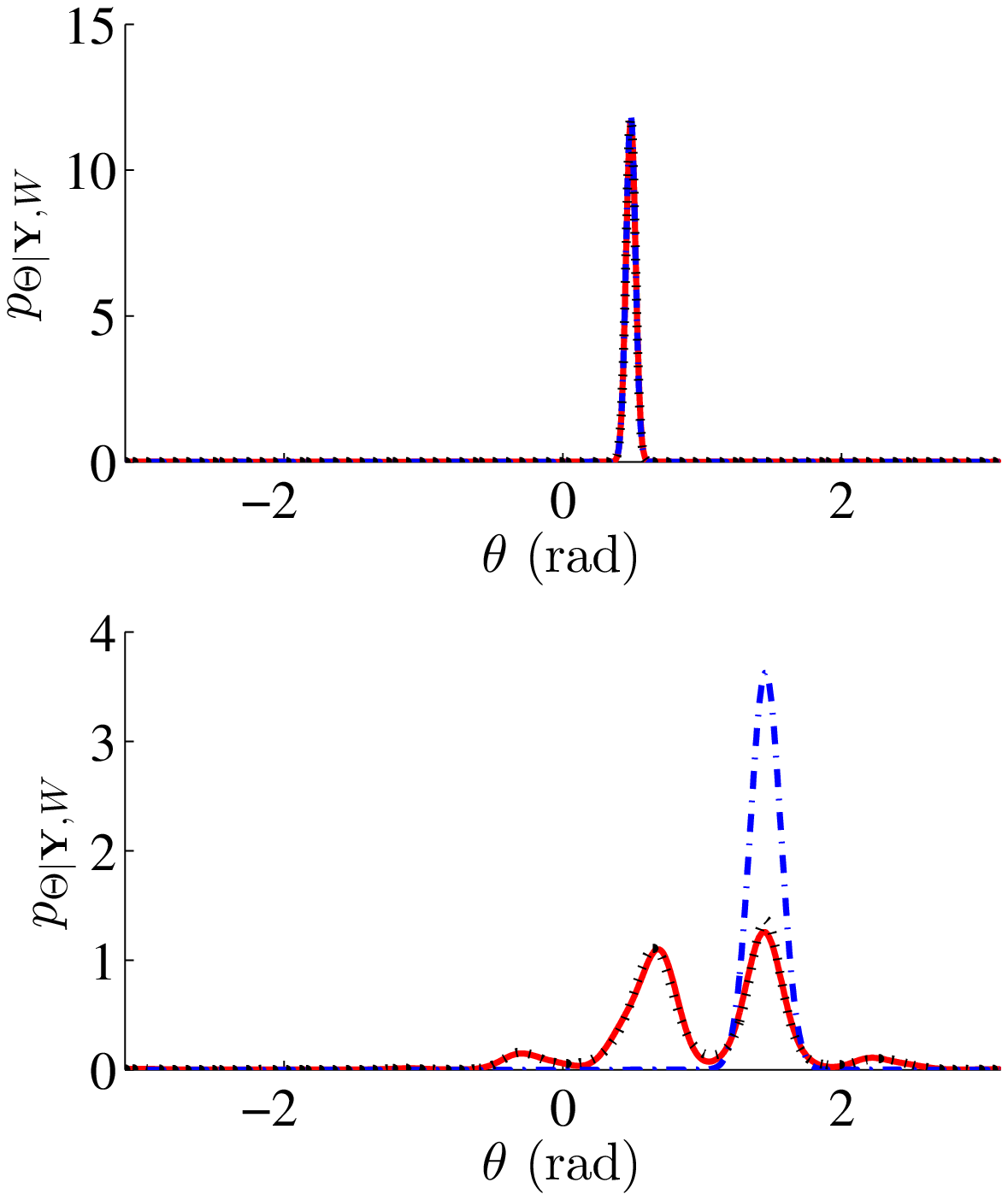}
    \label{fig:Example_coherent2}
}
\subfloat[]{
    \includegraphics[width=0.32\textwidth]{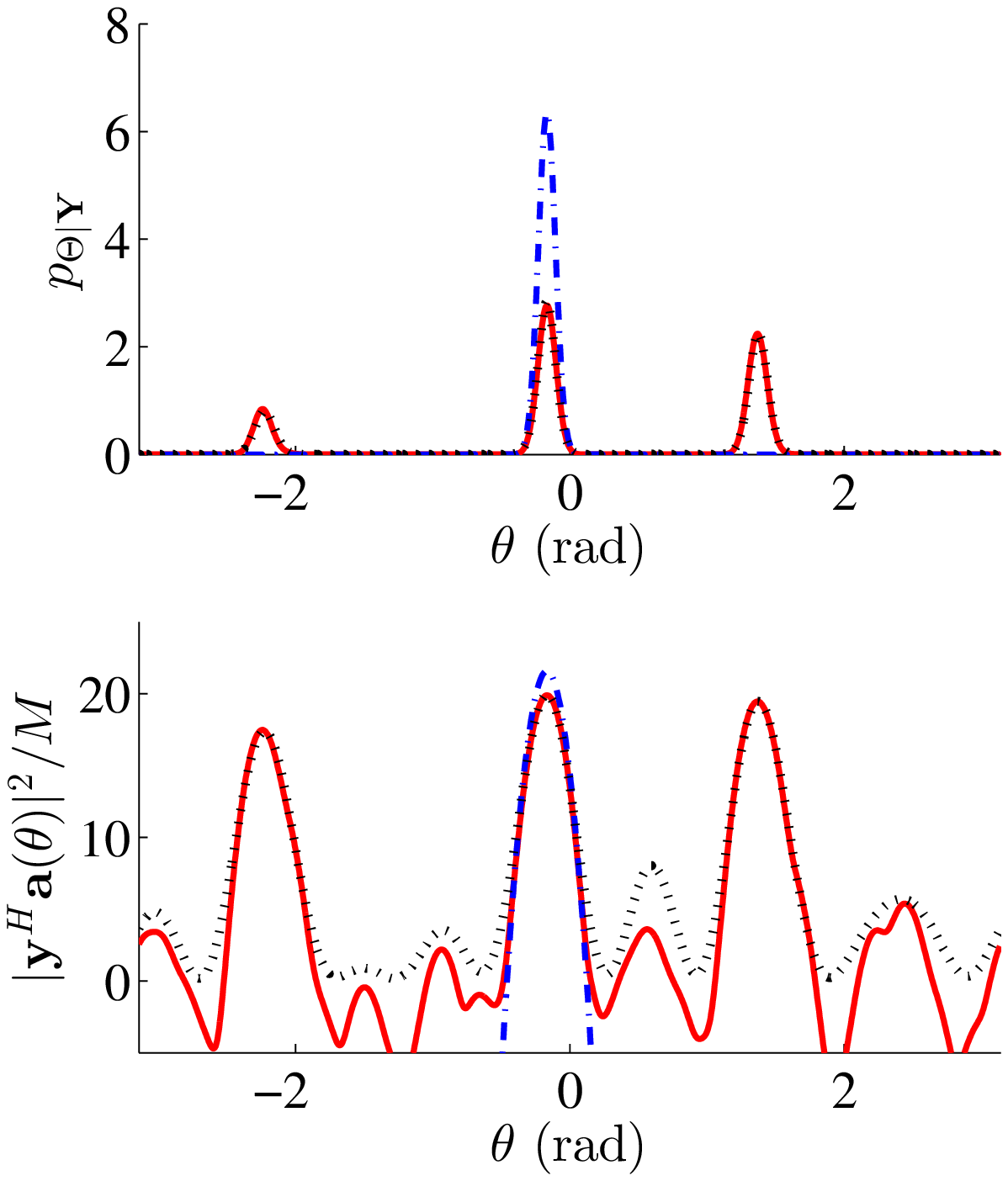}
    \label{fig:Example_noncoherent}
}
\caption{(a) Snapshot of the pdf~\eqref{eq:pdf_coh} and its approximations for $\theta = 0.5$, $\mathcal{M} = \{1,2\}$ and $\text{SNR}=0 \text{ dB}$ (top),  $\text{SNR}=10 \text{ dB}$ (bottom).
(b) Snapshot of the pdf~\eqref{eq:pdf_coh} and its approximations for $\theta = 0.5$, $\text{SNR}=0 \text{ dB}$ and $\mathcal{M} = \{1,\ldots,10\}$ (top), $\mathcal{M} = \{1,7,10\}$ (bottom).
(c) Snapshot of $p_{\Theta\mid\bm{Y}}$ in~\eqref{eq:pdf_noncoh} and its approximations for $K=3$, $\theta_1=-2.28$, $\theta_2=-0.04$, $\theta_3 = 1.39$, $\mathcal{M} = \{0,\ldots,9\}$, $\text{SNR}=3 \text{ dB}$ (top); using a log scale (bottom).
}
\label{fig:Ex_MVM}
\end{figure*}

Let us consider the estimation of the frequency of a single sinusoid when we know its weight $w$. In this case, the posterior pdf is
\begin{equation}\label{eq:pdf_coh}
    p_{\Theta\mid \bm{Y},W}(\theta\mid\vy,w)  \propto  \exp\left\{ \Re\left( \frac{2}{\nu}w\vy^{\Herm}\va(\theta) \right) \right\}.
\end{equation}
Fig.~\ref{fig:Example_coherent1} and~\ref{fig:Example_coherent2} display snapshots of the pdf~\eqref{eq:pdf_coh} and its approximations for different settings of $\mathcal{M}$ and $\text{SNR}$. When the number of measurements and SNR are low (Fig.~\ref{fig:Example_coherent1} top), the pdf is spread across its domain. The MVM approximation~\eqref{eq:blf_theta_MvM} has $1\times 2=2$ components; Heuristic~1 follows closely the exact pdf by keeping both components of the mixture, while the single VM pdf given by Heuristic~2 captures the dominant mode.
Increasing the SNR (Fig.~\ref{fig:Example_coherent1} bottom) makes the pdf more concentrated and both approximations are tight (in Heuristic~1, one component of the MVM has amplitude almost one and therefore the other is irrelevant). The pdf becomes more concentrated also by increasing the number of measurements, even though the SNR is low (Fig.~\ref{fig:Example_coherent2} top). Even though the approximation~\eqref{eq:blf_theta_MvM} has $10!$ components, among the $D=1000$ components output by Heuristic~1 only one is relevant. In the case of incomplete data (Fig.~\ref{fig:Example_coherent2} bottom), the pdf~\eqref{eq:pdf_coh} can have several significant modes. Among the $D=1\times 7\times 10=70$ components provided by Heuristic~1 (with this setting of $D$, all components in~\eqref{eq:blf_theta_MvM} are kept), only $12$ have amplitudes larger than $10^{-3}$. Heuristic~2 captures the largest mode and misses the mass containing the true $\theta$.

\subsubsection{Noncoherent estimation}
Without knowing the weight of the sinusoid, we can marginalize $p_{\Theta, W\mid\bm{Y}}$ and, assuming an improper ``flat'' prior of $W$, obtain
\begin{equation}
    p_{\Theta\mid\bm{Y}}(\theta\mid\bm{y}) = \int f_\text{CN} (\vy;w \va(\theta),\nu) \operatorname{d}\!w \propto \exp\left( \frac{|\vy^{\Herm} \va(\theta)|^2}{\nu M}  \right). \label{eq:pdf_noncoh}
\end{equation}
The exponent of~\eqref{eq:pdf_noncoh} is in fact the periodogram scaled by $1/\nu$. We write~\eqref{eq:pdf_noncoh} in a form favorable for the MVM approximation. First, let us define $\mathcal{M}^\prime = \{m-n\mid m,n\in\mathcal{M}, m>n \}$ with cardinality $M^\prime$ and the vector-valued function $\va^\prime:[-\pi,\pi)\rightarrow\mathbb{C}^{M^\prime}$, $\omega\to\va^\prime(\omega)\triangleq (e^{j\omega m} \mid m\in\mathcal{M^\prime})^{\operatorname{T}}$. By simply developing $|\vy^{\Herm} \va(\theta)|^2$ we arrive at
\begin{equation}\label{eq:noncoh_2}
    p_{\Theta\mid\bm{Y}}(\theta\mid\bm{y}) \propto \exp\left( \Re\left(\tfrac{2}{\nu}\bm{\gamma}^{\Herm} \va^\prime(\theta)\right) \right)
\end{equation}
where, for each $t=1,\ldots,M^\prime$, $\gamma_{t} = \frac{1}{M} \sum_{(k,l)\in\mathcal{T}_t} y_{k} y_{l}^\ast$ with $\mathcal{T}_t = \{(k,l) \mid 1\leq k,l \leq M, m_k-m_l=t\}$.\footnote{Actually, $\bm{\gamma}$ is the sample autocovariance of $\vy$.} Given~\eqref{eq:noncoh_2}, we can approximate $p_{\Theta\mid\bm{Y}}$ as an MVM~\eqref{eq:blf_theta_MvM}. In the log domain the approximation provides a representation of the periodogram.

As an illustration, we take $K=3$ and plot a snapshot of $p_{\Theta\mid\bm{Y}}$ (Fig.~\ref{fig:Example_noncoherent} top) and the log of $p_{\Theta\mid\bm{Y}}$ scaled so as to give the periodogram (Fig.~\ref{fig:Example_noncoherent} bottom). We can see again the good agreement between the approximations and the exact curves. The three lobes corresponding to each of the sinusoids are very well represented by Heuristic~1 while Heuristic~2 picks up the highest lobe. Due to the exponentiation, $p_{\Theta\mid\bm{Y}}$ is significant only at the values of $\theta$ for which $\va(\theta)$ is well aligned with $\vy$.

\section{The VALSE Algorithm}\label{sec:VALSE}
We define a schedule for iteratively updating the factors of $q_{\bm{\Theta},\bm{W},\bm{S}\mid\bm{Y}}$ and estimates $\hat\nu$, $\hat\rho$, $\hat\tau$ derived in Sec.~\ref{sec:SolVarOpt}, and propose an initialization of the scheme.\footnote{The alternating minimization scheme generates sequences of factors and estimates which, for notational convenience, we will not index with iteration numbers. It is to be understood that the update of one quantity depends on the most recent updates of the rest of the quantities.} The resulting algorithm, which we dub variational line spectral estimation (VALSE), is outlined in Algorithm~\ref{alg:VALSE}. Since each step increases the lower bound~\eqref{eq:free_energy}, the algorithm converges to some local maximum of $\mathcal{L}$. The stopping criterion can be defined in terms of the relative change of some quantity (e.g., $\mathbf{\hat x}$) from one iteration to the next or a maximum number of iterations.

\begin{algorithm}
\caption{Outline of the VALSE algorithm}
\label{alg:VALSE}
\begin{algorithmic}[1]
\renewcommand{\algorithmicrequire}{\textbf{Input:}}
\renewcommand{\algorithmicensure}{\textbf{Output:}}
\REQUIRE Signal vector $\vy$, set $\mathcal{M}$ of measurement indices
\ENSURE  Model order estimate $\hat K$, frequency and amplitude estimates $\{ (\hat\omega_k,\hat{\alpha}_k)\}_{k=1}^{\hat K}$, reconstructed signal $\mathbf{\hat{x}}$
\STATE{Initialize $\hat\nu$, $\hat\rho$, $\hat\tau$ and $q_{\Theta_i\mid\bm{Y}}$, $i\in\{1,\ldots,N\}$; compute $\hva_i$}
\REPEAT
    \STATE{Update $\hvs$, $\hvw_\mathcal{\hat{S}}$ and $\hmC_\mathcal{\hat{S}}$ (Algorithm~\ref{alg:maxZ})}
    \STATE{Update $\hat\nu$~\eqref{eq:est_noise_var}, $\hat\rho$ and $\hat\tau$~\eqref{eq:est_rho_tau}}
    \STATE{For all $i\in\hat{\mathcal{S}}$, update $\veta_i$~\eqref{eq:eta_1} and $\hva_i$ (Sec.~\ref{sec:approx_vonMises}) }
\UNTIL{stopping criterion}
\RETURN $\|\hvs\|_0$, $\hvtheta_{\hat{\mathcal{S}}}$, $\hvw_{\hat{\mathcal{S}}}$ and $\mathbf{\hat{x}}$~\eqref{eq:ReconSig}
\end{algorithmic}
\end{algorithm}

While several initialization schemes can be imagined, we choose to initialize $\{q_{\Theta_i\mid\bm{Y}}\}_{i=1}^N$ in a sequential manner. In the first step, we assign $q_{\Theta_1\mid\bm{Y}}$ the noncoherent pdf form~\eqref{eq:pdf_noncoh} and initialize the parameter estimates. For the latter, we use $\bm{\gamma}$ in~\eqref{eq:noncoh_2} (whose entries are estimates of the autocovariane function) to build a Toeplitz estimate of $\operatorname{E}[\vy\vy^{\Herm}]$. Then, we initialize $\hat{\nu}$ with the average of the lower quarter of the eigenvalues of that matrix. Given that $\operatorname{E}[\vy^{\Herm}\vy]/M = \rho N\tau + \nu$, we set $\hat{\rho} = 0.5$ and let $\hat{\tau} = (\vy^{\Herm}\vy/M - \hat{\nu}) /(\hat{\rho} N)$. Then, in step $i$, when we have initialized the first $i-1$ pdfs, we compute the estimates $\{\hat w_k\}_{k=1}^{i-1}$ and the residual $\vz_{i-1}=\vy - \sum_{k=1}^{i-1} \hat{w}_k \hva_k$. Initializing $q_{\Theta_i\mid\bm{Y}} \propto \exp\left\{ |\vz_{i-1}^{\Herm} \va(\theta)|^2 /(\nu M) \right\}$, we can represent $q_{\Theta_i\mid\bm{Y}}$ as an MVM (see~\eqref{eq:pdf_noncoh} and~\eqref{eq:noncoh_2}) and compute $\hva_i$.\footnote{In the initialization we use Heuristic~2 to compute the $\hva_i$'s because, when the sinusoidal components have similar powers, Heuristic~1 will capture contributions from the signal components that are not yet initialized, while Heuristic~2 picks up the strongest one (see Fig.~\ref{fig:Example_noncoherent}).}

The complexity per iteration is dominated by the maximization of $\ln Z(\vs)$ needed in line 3 (realized by Algorithm~\ref{alg:maxZ}) and the approximation of $\{q_{\Theta_i\mid\bm{Y}}\}_{i\in\hat{\mathcal{S}}}$ by mixtures of von Mises pdfs required in line 5 (using Algorithm~\ref{alg:Heuristic1} or Algorithm~\ref{alg:Heuristic2}). According to the analysis in Appendix~\ref{sec:approx_maxZ}, the maximization has complexity $\mathcal{O}(N\hat{K}^3)$ (actually, $\mathcal{O}(N\hat{K}^2)$ during most of the iterations of VALSE). As indicated in Sec.~\ref{sec:approx_vonMises}, the complexity of the MVM approximation is $\mathcal{O}(DMN)$ with Heuristic 1 and $\mathcal{O}(MN)$ with Heuristic 2; thus, the update of the pdfs of all frequencies with indices in $\hat{\mathcal{S}}$ has complexity $\mathcal{O}(\hat{K}DMN)$, respectively $\mathcal{O}(\hat{K}MN)$.

\section{Simulation Experiments}
In this section, we use computer simulations to assess the performance of the VALSE algorithm and state-of-the-art methods under different scenarios.

\subsection{Setup, metrics and algorithms}
Referring to~\eqref{eq:OrigSig}, the $K$ values $\{\omega_k\}_{k=1}^K$ of the angular frequencies are generated one-by-one: $\omega_k$ is drawn from $\mathcal{U}(-\pi,\pi)$ until a minimum (wrap-around) distance $\Delta\omega$ is ensured between $\omega_k$ and each of the $k-1$  previously generated values. The complex amplitudes $\{\alpha_k\}_{k=1}^K$ are generated randomly by drawing their magnitudes from $\mathcal{N}(1,0.1)$ and phases from $\mathcal{U}(-\pi,\pi)$. The noise samples contaminating the observations~\eqref{eq:SigModel} are independent and zero-mean complex Gaussian distributed.

The following metrics are evaluated by averaging from 500 independent trials: the normalized mean square error of signal reconstruction, $\operatorname{E}[\|\mathbf{\hat x}-\vx\|_2^2/\|\vx\|_2^2]$; the success rate, which we compute as the empirical probability of $\hat{K}=K$; and frequency estimation error. For a given simulation point, the frequency estimation error is evaluated only for the algorithms that provide a success rate $\geq 0.1$ by averaging only the trials in which all those algorithms output $\hat{K}=K$. The assignment of estimated components to the true ones is performed according to the Munkres' (or Hungarian) algorithm~\cite{Munkres1957} with the cost being the squared error of frequency estimates. We also report the runtime per trial for different problem sizes as an indicator of the complexity of the methods. The Cram\'{e}r-Rao lower bounds (CRLB) on the reconstruction and frequency estimation errors are computed by assuming $K$ is known.

We present the results for VALSE using Heuristic~2 to compute the estimates $\{\hva_i\}$ in line 5 of Algorithm~\ref{alg:VALSE}. In general, we obtain very similar performances with Heuristic~1 and Heuristic~2, the latter being significantly faster. Even though in the tough conditions of low SNR and/or few measurements Heuristic~1 provides better representation of the pdfs (see Fig.~\ref{fig:Ex_MVM}), we observed that in those conditions Heuristic~1 has the tendency to underestimate $K$ and provide slightly lower success rate than Heuristic~2. We assume no prior information about the frequencies is available, so we set $p_{\Theta_i}(\theta_i)=1/(2\pi)$, $i=1,\ldots,N$. Algorithm~\ref{alg:VALSE} stops at iteration $t$ if $\|\mathbf{\hat x}^{(t)}-\mathbf{\hat x}^{(t-1)}\|/\|\mathbf{\hat x}^{(t-1)}\| < 10^{-6}$ or the number of iterations reaches 5000.

We also introduce a variant of our algorithm, called ``VALSE-pt'', which operates with point estimates of the frequencies (as in the traditional approach). VALSE-pt additionally assumes that $q_{\Theta_i\mid\bm{Y}}(\theta_i\mid\vy)=\delta(\theta_i-\hat\theta_i)$ for all $i$, which gives that $\hat\theta_i$ is the maximizer of~\eqref{eq:blf_theta} and $\hva_i=\va(\hat\theta_i)$. We obtain $\hat\theta_i=\operatorname{arg\,max}\Re\left(\veta_i^{\Herm} \va(\theta_i)\right)$ numerically. Except for the computation of $\hva_i$ in line 5 of Algorithm~\ref{alg:VALSE}, all the other steps and settings of VALSE-pt and VALSE are identical.

For comparison, we evaluate the following state-of-the-art methods described in the Introduction: atomic-norm soft-thresholding\footnote{The software is available at \url{https://github.com/badrinarayan/astlinespec}. We used the implementation via ADMM.} (AST)~\cite{Bhaskar2013}---only applicable in the complete data case, the gridless-SPICE-based framework\footnote{The software was provided by the authors of~\cite{Yang2015}. We used the implementation via ADMM~\cite{Yang2015}.} (GLS)~\cite{Yang2015}, enhanced matrix completion\footnote{We used the software available at \url{http://www2.ece.ohio-state.edu/~chi/research.html}. The implementation uses the SDPT3 solver. Based on the ``cleaned'' signal output by EMaC, we perform model order and parameter estimation using Root-MUSIC and Akaike information criterion.} (EMaC)~\cite{Chen2014} and reweighted atomic-norm minimization\footnote{The software was provided by the authors of~\cite{Yang2016}. The implementation uses the SDPT3 solver.} (RAM)~\cite{Yang2016}. To configure algorithm-specific parameters, AST, EMaC and RAM require knowledge of the noise power. For each of these three methods we use the noise-variance estimation in~\cite{Bhaskar2013}, which computes $\hat{\nu}$ by averaging a lower part of the eigenvalues of an estimate of $\operatorname{E}[\vy \vy^{\Herm}]$. EMaC and RAM require an upper bound on the $\ell_2$ norm of the noise vector in order to search only among candidate solutions whose distances to the measurement $\vy$ are less than the bound; we set this bound to $\sqrt{(M+2\sqrt{M})\hat{\nu}}$, as suggested in~\cite{Yang2016}.

\subsection{Estimation from complete data}
\begin{figure*}
\centering
\subfloat[]{
    \includegraphics[width=0.32\textwidth]{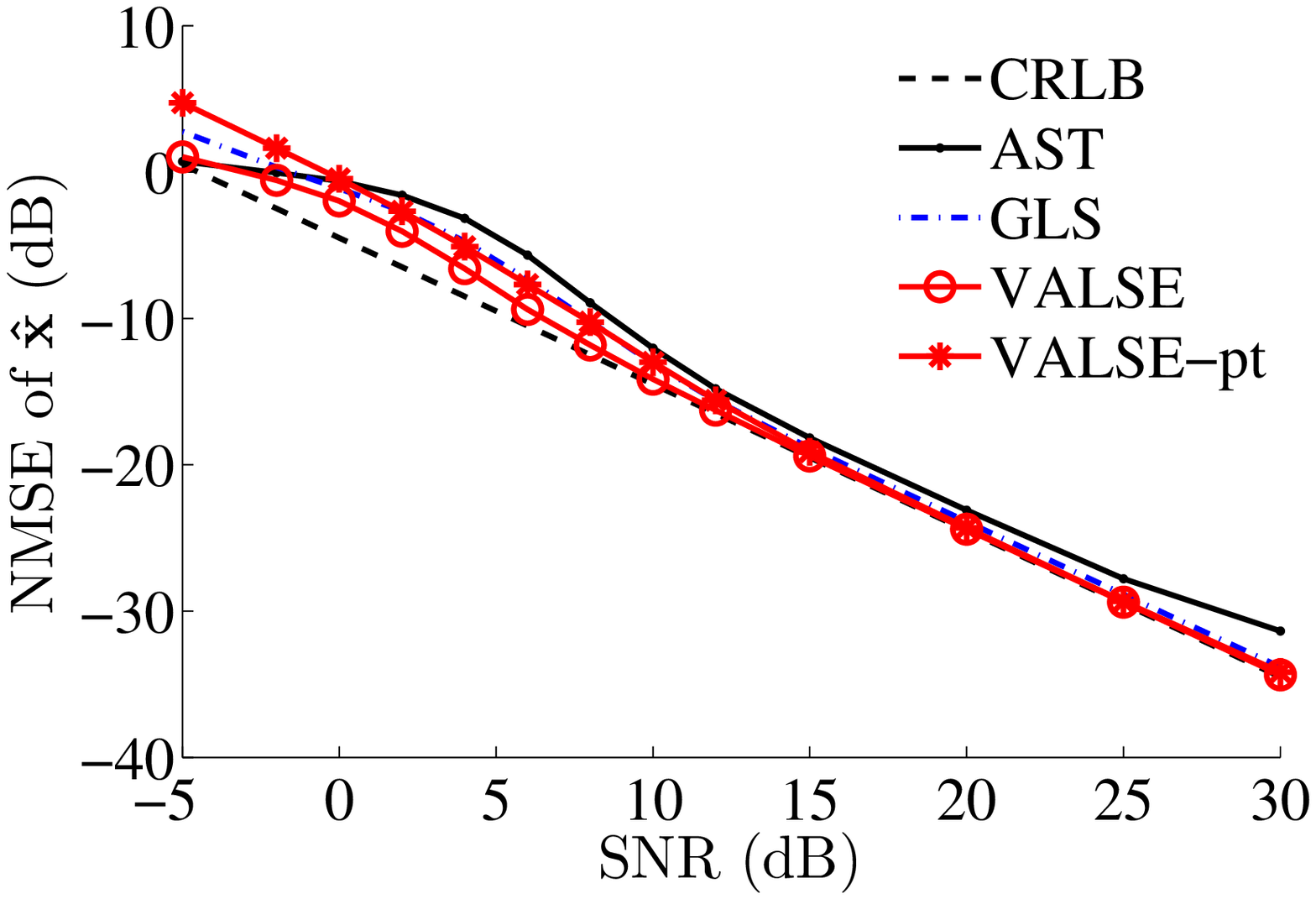}
    \label{fig:NMSE_SNR}
}
\subfloat[]{
    \includegraphics[width=0.32\textwidth]{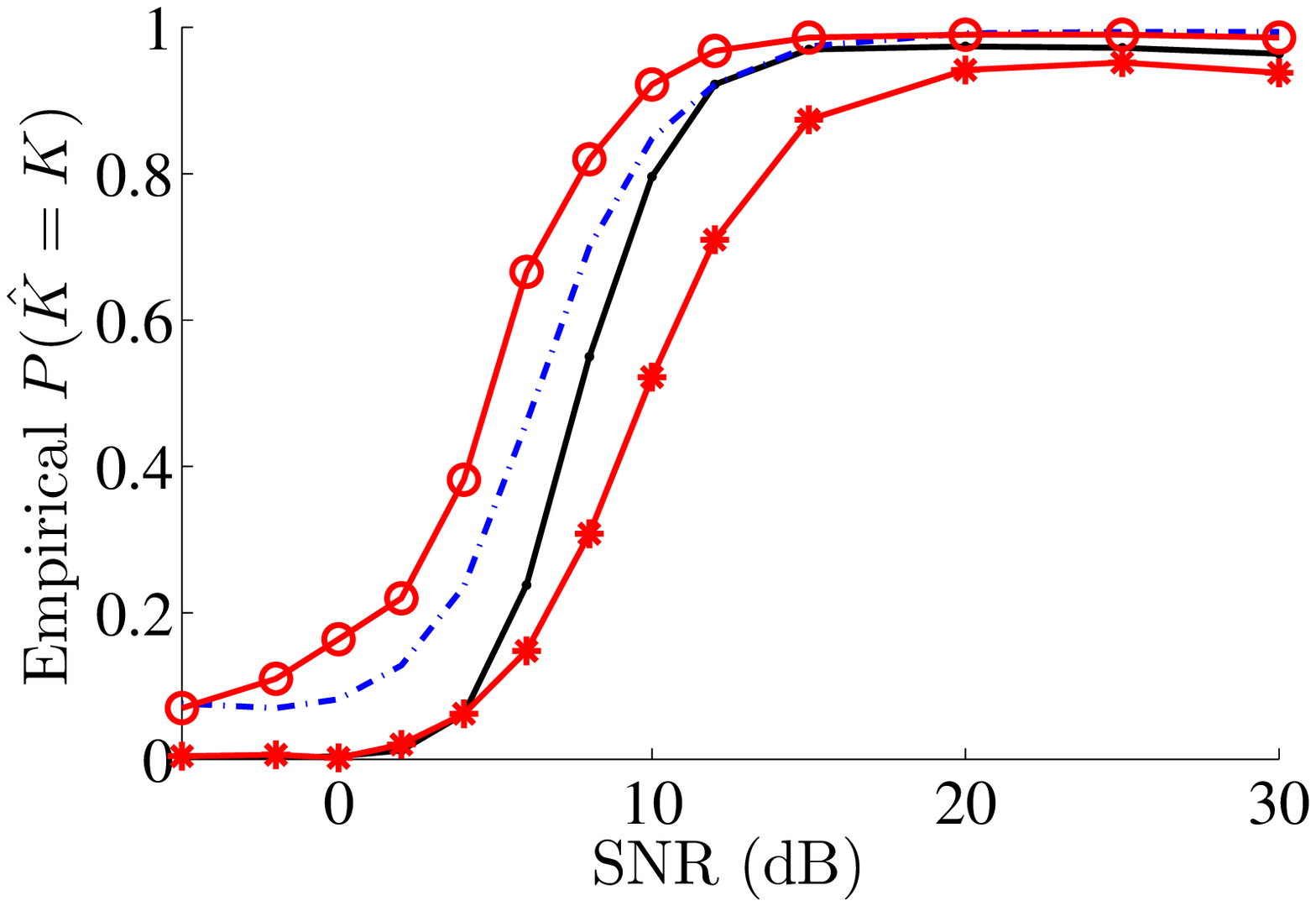}
    \label{fig:SR_SNR}
}
\subfloat[]{
    \includegraphics[width=0.32\textwidth]{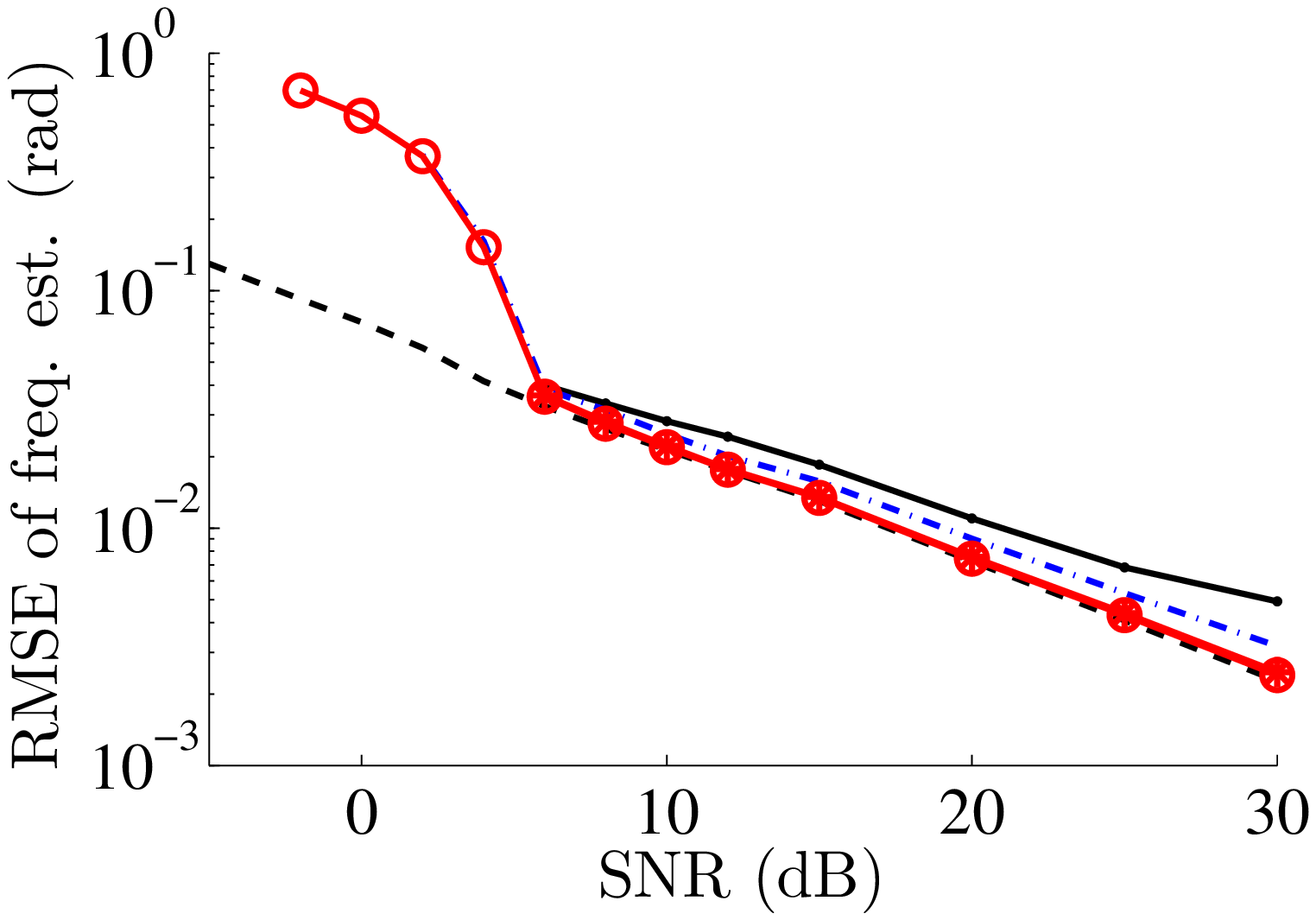}
    \label{fig:FreqMSE_SNR}
}
\caption{Performance vs. SNR for $M=N=21$ samples, $K=5$ and minimum separation $\Delta\omega=\frac{2\pi}{N}$: (a) normalized MSE of the reconstructed signal; (b) success rate of model order estimation; (c) root MSE for frequency estimation.}
\label{fig:Results_vsSNR}
\end{figure*}

Fig.~\ref{fig:Results_vsSNR} displays the results of estimating $K=5$ sinusoidal components from $M=N=21$ measurements at different SNR values. The distance between any two frequencies is at least $\Delta\omega=\frac{2\pi}{N}$ radians. VALSE outperforms the reference methods at all SNR values and shows excellent performance at $\text{SNR}\geq 10\text{ dB}$, where the reconstruction and frequency estimation errors are very close to the CRLB and the success rate is almost one. AST and GLS estimate the model order accurately as well in high SNR, but their success rates decrease earlier. The success rate of AST seems to saturate at a value slightly below one and degrades faster than that of GLS when the SNR decreases. In Fig.~\ref{fig:NMSE_SNR}, the gap between AST and VALSE increases for $\text{SNR}\geq 10\text{ dB}$, while GLS maintains a constant gap of about $0.5\text{ dB}$. A similar behavior can be also observed for the frequency estimation error in Fig.~\ref{fig:FreqMSE_SNR}.

We have also evaluated the EMaC and RAM algorithms. Since they did not provide significant improvements over AST and for the clarity of the figures, we do not show those results. In fact, we observed that EMaC and RAM perform well only at high SNR (above $20$-$25$ dB) where their success rates approach one; still, in this SNR regime, EMaC shows worse signal reconstruction and frequency estimation than AST, while RAM provides slight improvement over AST. For $\text{SNR}< 20\text{ dB}$, both EMaC and RAM are outperformed by AST in all metrics. Our explanation for their not so good performance in the low-to-moderate SNR region is that, according to our observations, their performance is quite sensitive to the setting of the upper bound on the $\ell_2$ norm of the noise vector and therefore to the accuracy of the noise variance estimate.

The gap between the success rates of VALSE-pt and VALSE is due to the former's tendency to overestimate $K$ more heavily. For example, at $\text{SNR} = 15 \text{ dB}$, VALSE outputs $\hat{K}=6$ in $5$ out of the $500$ simulation trials, while VALSE-pt outputs $\hat{K}=6$, $7$ and $10$ components in $53$, $7$ and respectively $1$ trials. The discrepancy between their performance comes from the way in which $\hva_i$ is computed in line 5 of Algorithm~\ref{alg:VALSE}, since this is the only difference between the two algorithms. VALSE computes $\hva_i=\operatorname{E}_{q_{\Theta_i\mid\bm{Y}}}[\va(\Theta_i)]$, which involves the expectations of the phasors $e^{jn\Theta_i}$. The more concentrated $q_{\Theta_i\mid\bm{Y}}$, the closer $|\operatorname{E}_{q_{\Theta_i\mid\bm{Y}}}[e^{jn\Theta_i}]|$ is to one and $\|\hva_i\|_2$ to $\sqrt{M}$. Therefore, the uncertainty in frequency estimation captured by $q_{\Theta_i\mid\bm{Y}}$ is reflected in $\hva_i$. Consequently, the uncertainty impacts all the other estimates, which in turn determine the component-acceptance criterion, and therefore influences the model order estimate. On contrary, VALSE-pt assigns $\hva_i=\va(\hat\theta_i)$ and thus puts full certainty on the phasors' estimates. Loosely speaking, VALSE-pt might include excessive components because it ``overtrusts'' them---this is what we also observe experimentally.

\subsection{Estimation from incomplete data}
\begin{figure*}
\centering
\subfloat[]{
    \includegraphics[width=0.32\textwidth]{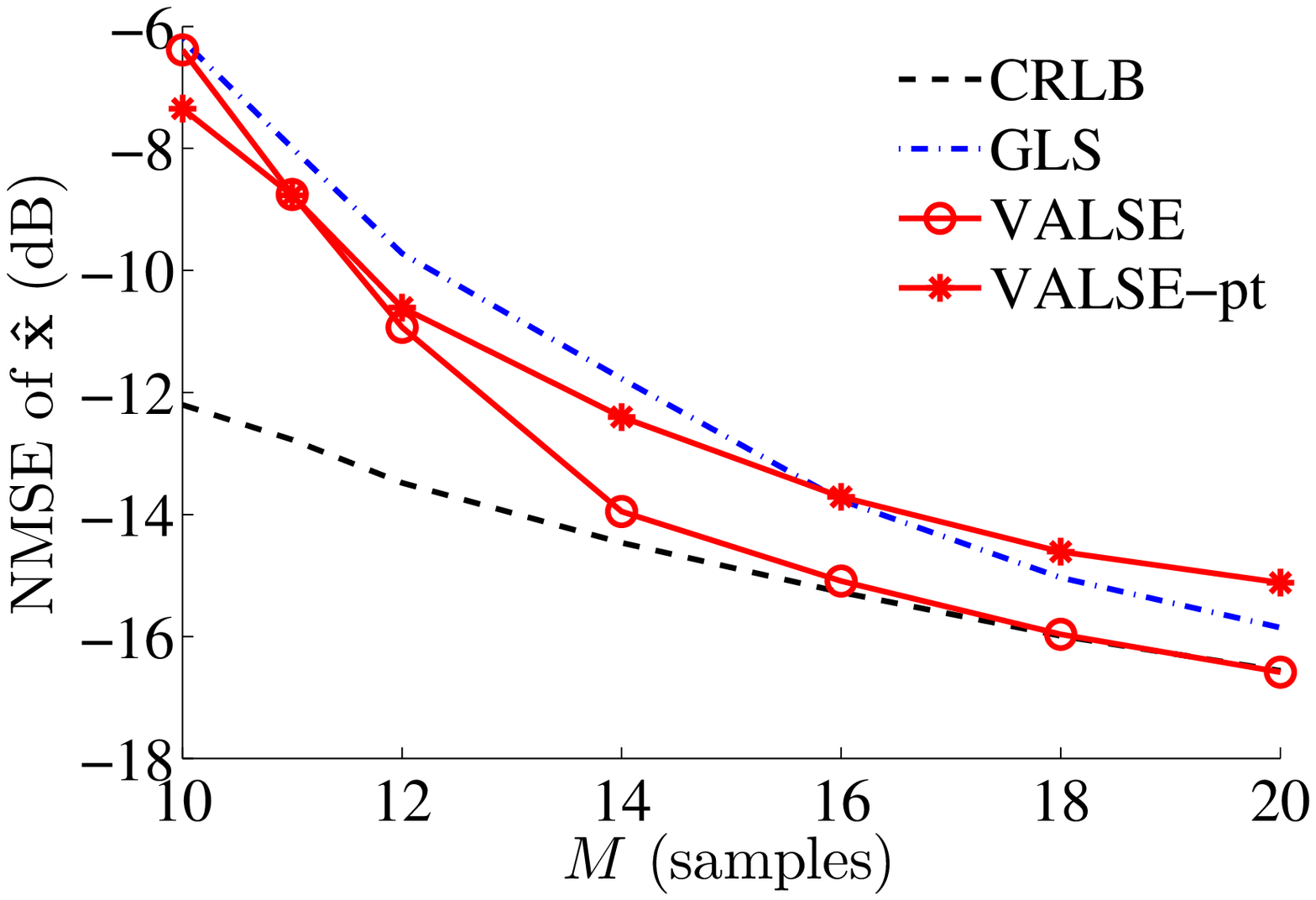}
    \label{fig:NMSE_M}
}
\subfloat[]{
    \includegraphics[width=0.32\textwidth]{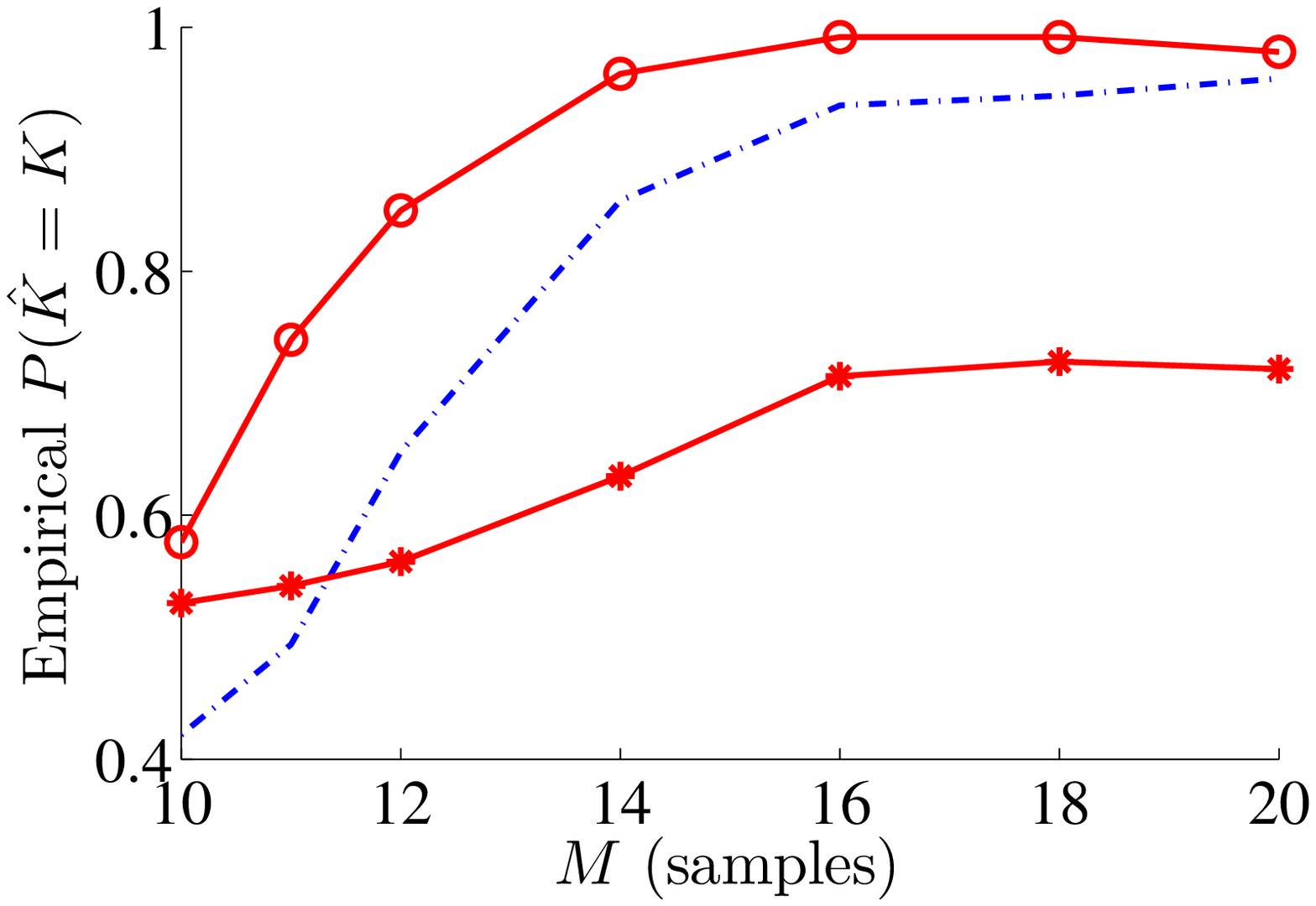}
    \label{fig:SR_M}
}
\subfloat[]{
    \includegraphics[width=0.32\textwidth]{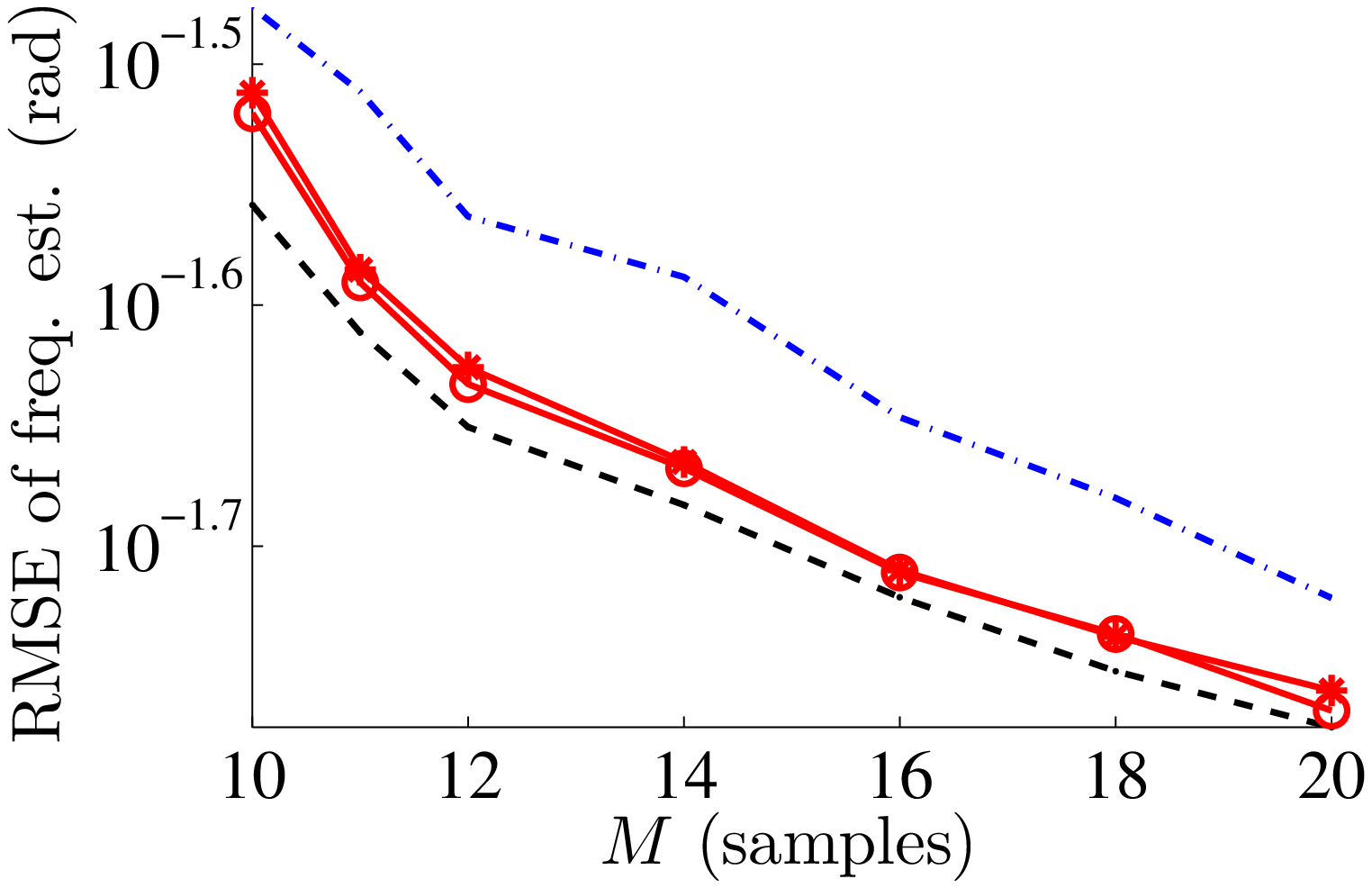}
    \label{fig:FreqMSE_M}
}
\caption{Performance vs. $M$  for $K=3$, $N=20$ samples, $\text{SNR}=10 \text{ dB}$ and minimum separation $\Delta\omega=\frac{2\pi}{N}$: (a) normalized MSE of the reconstructed signal; (b) success rate of model order estimation; (c) root MSE for frequency estimation.}
\label{fig:Results_vsM}
\end{figure*}

We now study the performance when the measurement data is incomplete, i.e., $M<N$. We consider the estimation of $K=3$ sinusoids when $N=20$ and $\text{SNR}=10 \text{ dB}$. The frequencies are separated by at least $\Delta\omega=\frac{2\pi}{N}$. Based on the previous analysis, in the comparison we include only the GLS method. The results in Fig.~\ref{fig:NMSE_M} and~\ref{fig:SR_M} show that, for $M\geq 14$, VALSE estimates $\vx$ very accurately (close to the CRLB) and selects the correct model order with a rate close to one. On contrary, the reconstruction errors of GLS and VALSE-pt are $1$--$2$ dB larger in that range of $M$. GLS provides a good estimation of $K$, although the success rate is always lower than that of VALSE and decreases earlier when reducing $M$. VALSE-pt shows a significantly lower success rate (again due to overestimation). When the algorithms estimate $K$ correctly, both VALSE and VALSE-pt provide very accurate frequency estimation, while GLS has larger errors.

\subsection{Resolution capability}
\begin{figure*}
\centering
\centering
\subfloat[]{
    \includegraphics[width=0.32\textwidth]{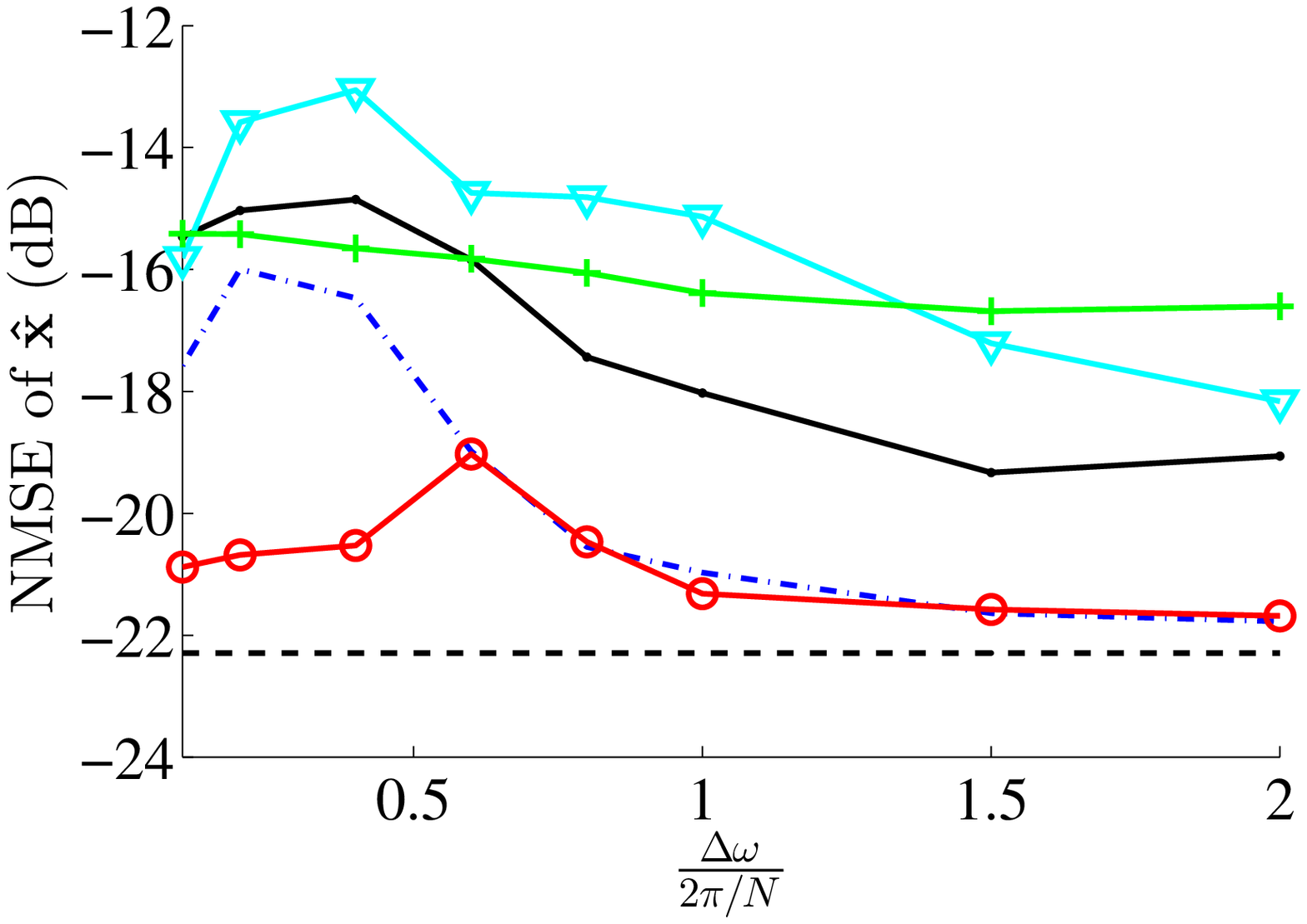}
    \label{fig:NMSE_D}
}
\subfloat[]{
    \includegraphics[width=0.32\textwidth]{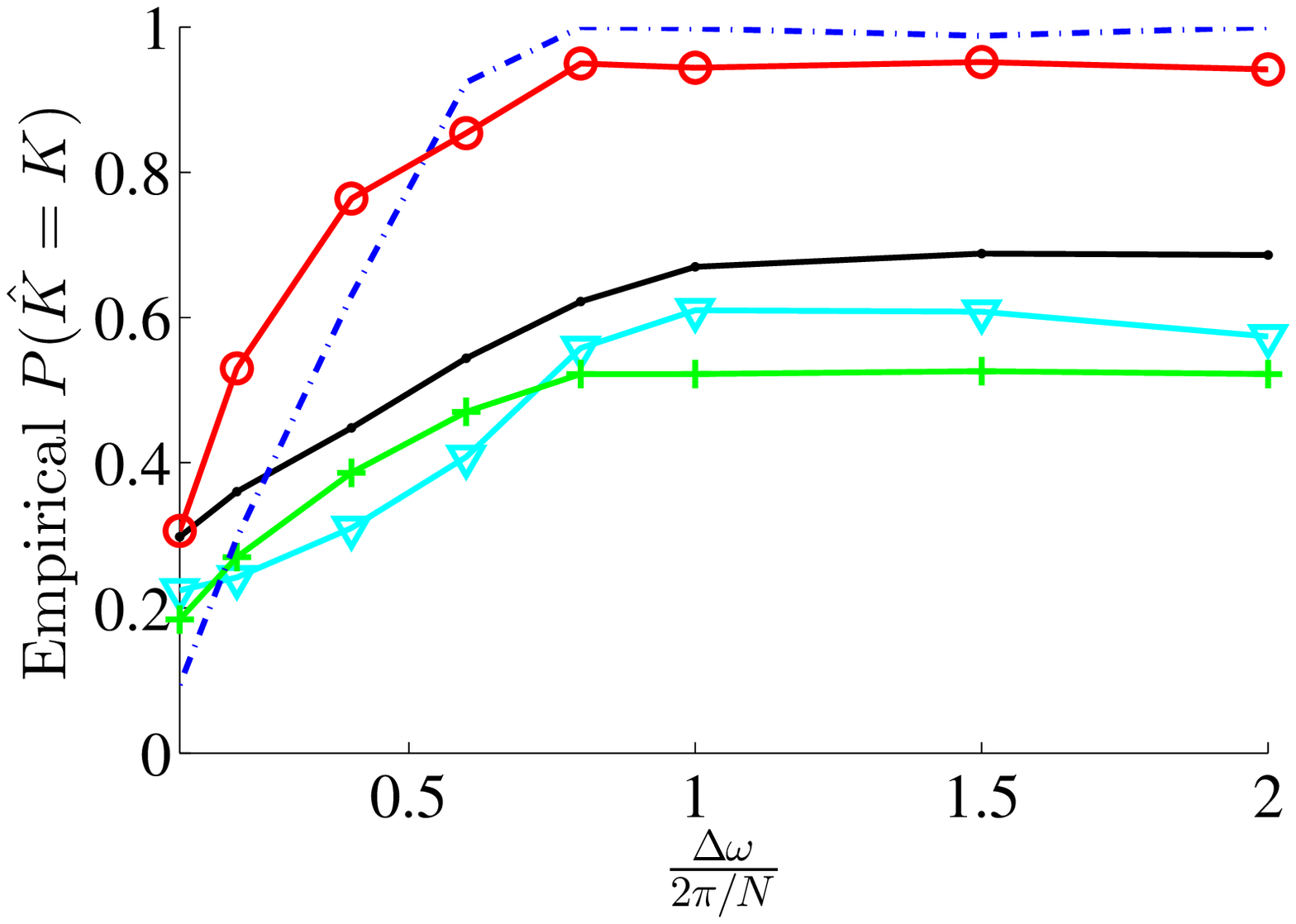}
    \label{fig:SR_D}
}
\subfloat[]{
    \includegraphics[width=0.32\textwidth]{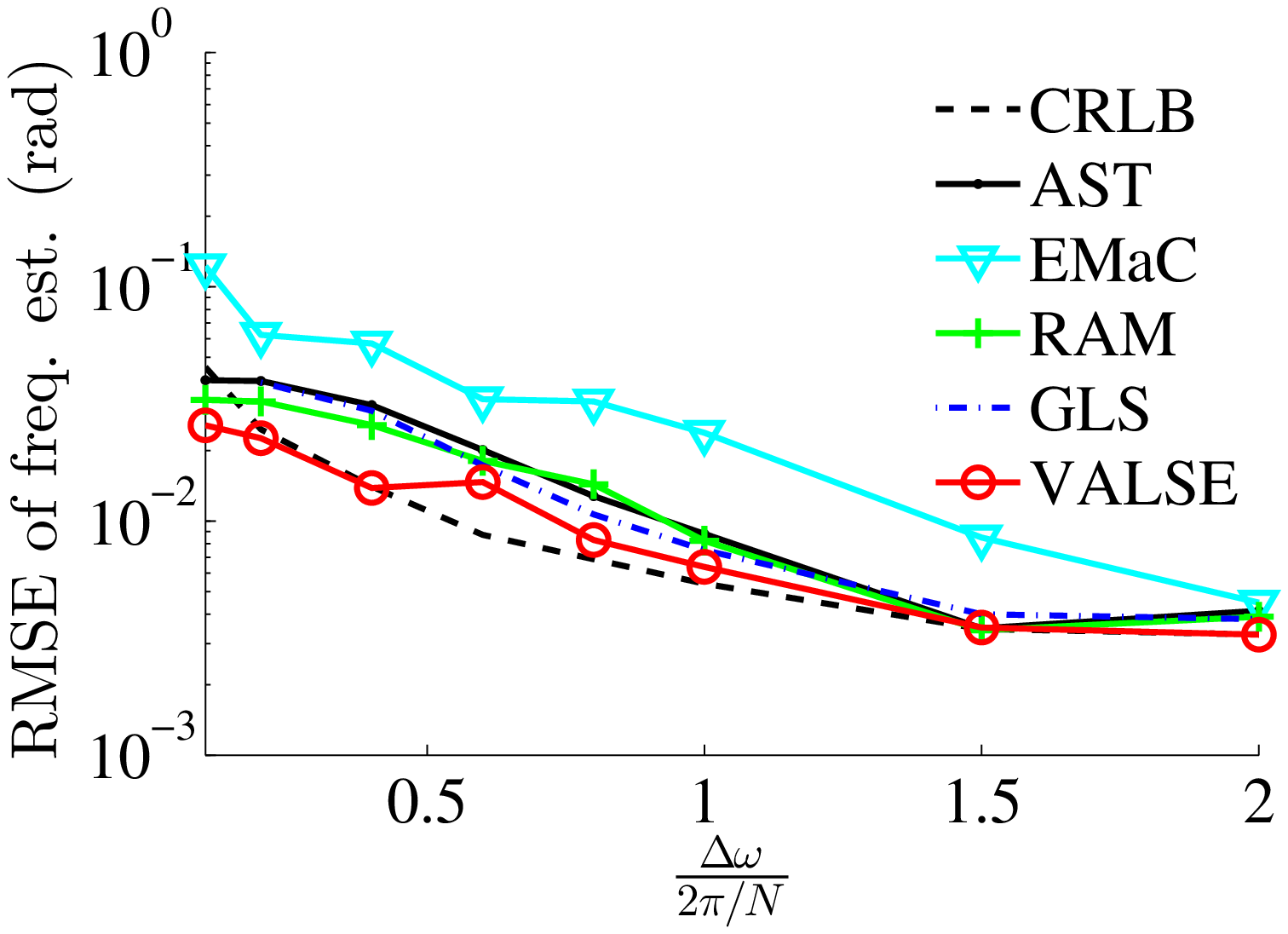}
    \label{fig:FreqMSE_D}
}
\caption{Performance of resolving $K=2$ sinusoids separated by small $\Delta\omega$; $M=N=51$ samples, $\text{SNR}=10 \text{ dB}$ and $0.1\times\frac{2\pi}{N}\leq \Delta\omega\leq 2\times\frac{2\pi}{N}$: (a) normalized MSE of the reconstructed signal; (b) success rate of model order estimation; (c) root MSE for frequency estimation.}
\label{fig:Results_K2}
\end{figure*}

Next, we evaluate the performance of resolving $K=2$ sinusoids that are closely-spaced in frequency. We draw $\omega_1$ from $\mathcal{U}(-\pi,\pi)$ and set $\omega_2 = \omega_1 + \Delta\omega$ (i.e., we impose an exact separation of $\Delta\omega$, and not a minimum one as in the previous experiments). Fig.~\ref{fig:Results_K2} shows results for $M=N=51$, $\text{SNR}=10 \text{ dB}$ and $0.1\times\frac{2\pi}{N}\leq \Delta\omega\leq 2\times\frac{2\pi}{N}$. We observe that, for $\Delta\omega > 0.5\times\frac{2\pi}{N}$, VALSE and GLS reconstruct the signal similarly well and estimate the correct model order with high probability (the success rate of VALSE seems to cap at about 0.95 while that of GLS comes very close to 1). When the two frequencies are separated by less than $0.5\times\frac{2\pi}{N}$, VALSE shows a significantly higher success rate compared to GLS; the reconstruction performance of the latter also degrades considerably. Fig.~\ref{fig:FreqMSE_D} shows that VALSE estimates the frequencies accurately in the whole range of $\Delta\omega$. AST, EMaC and RAM provide significantly lower performance in reconstructing the signal and selecting the model order, which is inline with our observations in the first experiment.

\subsection{Scaling with the problem size}
\begin{figure}
\centering
\includegraphics[width=0.8\columnwidth]{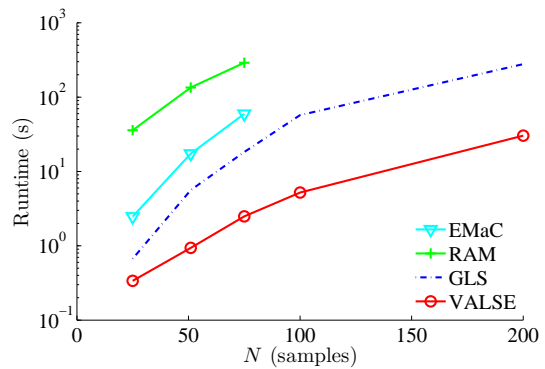}
\caption{Scaling of the runtime with the problem size. The simulation points correspond to the following $(N,M,K)$ triples: $(25,15,2)$, $(51,30,4)$, $(75,45,6)$, $(100,60,8)$ and $(200,120,16)$.}
\label{fig:Runtime}
\end{figure}

To obtain an indication of how the complexity of VALSE scales with the dimension of the problem, we evaluate the runtime for different sizes $N$. We consider an incomplete-data scenario in which the number $M$ of measurements and model order $K$ scale with $N$ and $\text{SNR}=20 \text{ dB}$. The following $(N,M,K)$ triples are investigated: $(25,15,2)$, $(51,30,4)$, $(75,45,6)$, $(100,60,8)$ and $(200,120,16)$. The results in Fig.~\ref{fig:Runtime} clearly show that VALSE is computationally advantageous compared to the benchmark methods. While RAM's runtime becomes quickly prohibitive, followed by EMaC and GLS, VALSE is about $10$ times faster than GLS when $N$ increases.

\section{Conclusions}
In this paper, we treated line spectral estimation (LSE) as Bayesian inference in a probabilistic model of the frequencies and coefficients. The latter were modeled by a Bernoulli-Gaussian distribution, which turned model order selection into detection of a binary sequence. To circumvent the deadlock of exact inference we resorted to the variational approach in which an approximate (surrogate) posterior pdf was computed analytically by maximizing a lower bound on the model evidence. Contrary to related works which compute point estimates of the frequencies, we considered estimating and working with their posterior probability density functions (pdfs). We showed that these pdfs can be very well approximated by mixtures of von Mises pdfs, which enables computation of closed-form expectations. In fact, our simulations show that the representation by one von Mises pdf seems appropriate.  The resulting VALSE algorithm increases the lower bound on the model evidence in each step and hence is convergent. Since all the parameters are estimated, VALSE does not require any fine tuning by the user. Simulation results advocate our fully Bayesian approach of representing and operating with the uncertainty in frequency estimation, as we obtain significantly improved performance compared to just using point estimates. VALSE shows an excellent performance (often close to the Cram\'{e}r-Rao bound), consistently better than the benchmark.

Our method can straightforwardly include prior knowledge about the frequencies in the form of von Mises pdfs or mixtures of such pdfs if multimodal distribution are desired. The fact that VALSE conveniently represents posterior distributions allows for estimating the uncertainty in the estimation. Also, the pdfs can be subsequently used as prior pdfs in applications that rely on line spectral estimation. As an outlook, we expect that finding a better variational approximation, in which the surrogate pdf does not fully factorize over the frequencies yet still facilitates tractable inference, would further improve performance, especially in situations where the frequencies are very closely spaced.

\appendices
\section{Finding a local maximum of $\ln Z(\vs)$} \label{sec:approx_maxZ}
To find the globally optimal binary sequence $\vs$ would require $2^N$ evaluations of $\ln Z(\vs)$ given by~\eqref{eq:Z}. Inspired by the iterative search strategy proposed in~\cite{Kormylo1982}, we seek a locally optimal solution in a progressive manner. In step $p$, the utility of the reference sequence $\vs^{(p)}$ is compared to the $N$ utilities corresponding to $N$ test sequences. Specifically, the $k$th test sequence $\mathbf{t}_k$ is obtained by flipping the $k$th location of $\vs^{(p)}$. The change $\Delta_k^{(p)} = \ln Z(\mathbf{t}_k) - \ln Z(\vs^{(p)})$ is evaluated for each $k=1,\ldots,N$, and the test sequence giving the highest positive change is used as the reference sequence $\vs^{(p+1)}$ in the next step. If $\Delta_k^{(p)}<0$ for all $k=1,\ldots,N$, then the search stops and we set $\hvs = \vs^{(p)}$. The search starts with a certain initial reference sequence $\vs^{(0)}$ and converges in a finite number of steps to some locally optimal sequence. Although~\eqref{eq:Z} involves a matrix inversion, the changes $\Delta_k^{(p)}$, $k=1,\ldots,N$, can be efficiently computed in each step $p$ as follows.

Assume we change a sequence $\vs$ into a sequence $\vs^\prime$ by flipping the bit at the $k$th location. When $k\notin\mathcal{S}$, i.e. $s_k=0$, $s^\prime_k=1$ and $\mathcal{S^\prime}=\mathcal{S}\cup\{k\}$, we say the $k$th location is activated. Using the formulas for block-matrix determinant and block-wise matrix inversion, we write $\ln Z(\vs^\prime) - \ln Z(\vs)$ as
\begin{equation}\label{eq:DZactiv}
    \Delta_k = \ln\frac{v_k}{\tau} + \frac{|u_k|^2}{v_k} + \ln\frac{\rho}{1-\rho}.
\end{equation}
where
\begin{equation}\label{eq:uv}
\begin{split}
    v_k &= \nu\left(M+\frac{\nu}{\tau} - \nu^{-1} \mathbf{j}_k^{\Herm} \hmC_{\mathcal{S}} \mathbf{j}_k\right)^{-1} \\
    u_k &= \nu^{-1}v_k (h_k - \mathbf{j}_k^{\Herm} \hvw_{\mathcal{S}})
\end{split}
\end{equation}
with $\mathbf{j}_k = (J_{ik}\mid i\in\mathcal{S})^{\Tran}$. Upon changing $\vs$ into $\vs^\prime$, we use rank-one updates for the mean and covariance of the weights
\begin{equation}\label{eq:w_activ}
    \hat w^\prime_i =
    \left\{
    \begin{array}{ll}
        u_k, & \quad i = k, \\
        \hat w_i - \mathbf{\hat c}_i^{\Herm}\mathbf{j}_k u_k , & \quad i\in\mathcal{S}.
    \end{array}
    \right.
\end{equation}
and
\begin{equation}\label{eq:C_activ}
    \left(
       \begin{array}{cc}
         \hmC^\prime_{\mathcal{S}} & \mathbf{\hat c}^\prime_k \\
         \mathbf{\hat c}^{\prime^{\Herm}}_k & \hat C^\prime_{kk} \\
       \end{array}
     \right)
     =
     \left(
       \begin{array}{cc}
         \hmC_{\mathcal{S}} & \mathbf{0} \\
         \mathbf{0} & 0 \\
       \end{array}
     \right)
     + v_k
     \left(
       \begin{array}{c}
         \hmC_{\mathcal{S}} \mathbf{j}_k\\
         -1 \\
       \end{array}
     \right)
     \left(
       \begin{array}{c}
         \hmC_{\mathcal{S}} \mathbf{j}_k\\
         -1 \\
       \end{array}
     \right)^{\Herm}
\end{equation}
Thus, by activating the $k$th component, the posterior mean and variance of $W_k$ are $u_k$ and $v_k$, respectively.

In the case of deactivation, i.e., $s_k=1$, $s_k^\prime=0$ and $\mathcal{S}^\prime=\mathcal{S}\setminus\{k\}$, the change $\ln Z(\vs^\prime) - \ln Z(\vs)$ is given by
\begin{equation}\label{eq:DZdeact}
    \Delta_k =- \ln\frac{\hat C_{kk}}{\tau} - \frac{|\hat w_k|^2}{\hat C_{kk}} - \ln\frac{\rho}{1-\rho}.
\end{equation}
We can again develop efficient updates: for all $i,j\in\mathcal{S}^\prime$,
\begin{equation}\label{eq:w_C_deact}
    \hat w_i^\prime = \hat w_i - \frac{\hat C_{ik}}{\hat C_{kk}}\hat{w}_k \quad\text{and}\quad
    \hat C^\prime_{ij} = \hat C_{ij} - \frac{\hat C_{ik} \hat C_{kj}}{\hat C_{kk}}.
\end{equation}

The iterative maximization is given by Algorithm~\ref{alg:maxZ}. The most expensive computation is to obtain $u_k$ for all $k\notin\mathcal{S}$ (in line~3). It requires $\mathcal{O}((N-l)l^2)$ operations, where $l=\|\vs\|_0$ is the current number of active locations. If in line $1$ we initialize $\vs=\mathbf{0}$ (i.e., $l=0$), the algorithm will execute the while loop $\mathcal{O}(\hat{K})$ times to output $\hvs$, where $\hat{K}=\|\hvs\|_0$. This gives the overall complexity $\mathcal{O}(N\hat{K}^3)$. However, in line $1$ we can initialize $\vs$ with $\hvs$ from the previous iteration of VALSE (Algorithm~\ref{alg:VALSE}). In this case, we observed that in each iteration of VALSE (except for the first one), the number of locations of $\hvs$ that are changed by Algorithm~\ref{alg:maxZ} is very small (in fact, often zero!). Thus, empirically, the complexity of Algorithm~\ref{alg:maxZ} during most of the iterations of VALSE is $\mathcal{O}(N\hat{K}^2)$.

\begin{algorithm}
\caption{Algorithm for maximizing $\ln Z(\vs)$}
\label{alg:maxZ}
\begin{algorithmic}[1]
\renewcommand{\algorithmicrequire}{\textbf{Input:}}
\renewcommand{\algorithmicensure}{\textbf{Output:}}
\REQUIRE $\mJ$, $\vh$, $\nu$ and $\rho$
\ENSURE  $\hvs$, $\hvw_\mathcal{\hat{S}}$ and $\hmC_\mathcal{\hat{S}}$
\STATE{Initialize $\vs$ and compute $\hvw_{\mathcal{S}}$ and $\hmC_{\mathcal{S}}~\eqref{eq:W_mean_var}$}
\WHILE{true}
    \STATE{For each $k\notin\mathcal{S}$, compute $u_k$ and $v_k$~\eqref{eq:uv}, and $\Delta_k$~\eqref{eq:DZactiv}}
    \STATE{For each $k\in\mathcal{S}$, compute $\Delta_k$~\eqref{eq:DZdeact}}
    \IF{$\{k\mid\Delta_k>0\} \neq\emptyset$}
        \STATE{$k_\ast= \operatorname*{arg\,max}_{k} \Delta_k$}
        \STATE{If $s_{k_\ast}=0$ compute ~\eqref{eq:w_activ}~\eqref{eq:C_activ}, else compute~\eqref{eq:w_C_deact}}
        \STATE{$s_{k_\ast} \gets s_{k_\ast}\oplus 1$}
    \ELSE
        \STATE{\textbf{break}}
    \ENDIF
\ENDWHILE\\
\RETURN $\hvs=\vs$, $\hvw_\mathcal{\hat{S}}=\hvw_\mathcal{S}$ and $\hmC_\mathcal{\hat{S}}=\hmC_\mathcal{S}$
\end{algorithmic}
\end{algorithm}

\section{Approximation of wrapped von Mises distributions}\label{app:approx}
The $N$-fold wrapped VM distribution is invariant under the transformation $\Theta\mapsto \Theta+\tfrac{2\pi}{N}$~\cite[p. 52]{MardiaJupp2000}. Its pdf is $f_\text{VM}(N\theta;\eta)$, for some $\eta = \kappa e^{j N \mu}$. The $N$ modes of the pdf have equal amplitudes and are evenly distributed around the circle, i.e., they are at $\mu + 2\pi n/N$, $n=0,\ldots,N-1$. We show that such a distribution is well approximated by an appropriate mixture of von Mises distributions (MVM) obtained by matching their characteristic functions. Our result extends the one in~\cite[p. 54]{MardiaJupp2000} which proposes the approximation for $N=2$.

The characteristic function $\varphi_p'$, $p\in\mathbb{Z}$, of a random variable having an $N$-fold wrapped VM distribution is
\begin{align}
    \varphi_p' &= \int_0^{2\pi} e^{jp\theta} \frac{1}{2\pi I_0(\kappa)} e^{\kappa\cos N(\theta-\mu)} \operatorname{d}\!\theta \nonumber \\
    &= \frac{e^{jp\mu}}{2\pi I_0(\kappa)} \sum_{n=0}^{N-1}\int_{2\pi n/N}^{2\pi(n+1)/N} e^{jp\theta} \, e^{\kappa\cos N\theta} \operatorname{d}\!\theta \nonumber \\
    &= \frac{e^{jp\mu}}{2\pi I_0(\kappa)} \sum_{n=0}^{N-1} e^{j2\pi \frac{p}{N} n} \int_0^{2\pi/N} e^{jp\theta} \, e^{\kappa\cos N\theta} \operatorname{d}\!\theta \nonumber \\
    &= \frac{e^{jp\mu}}{2\pi I_0(\kappa)} \frac{1}{N} \sum_{n=0}^{N-1} e^{j2\pi \frac{p}{N} n} \int_0^{2\pi} e^{j\frac{p}{N}\theta} \, e^{\kappa\cos\theta} \operatorname{d}\!\theta \label{eq:charfunc_wrap_0}
\end{align}
The sum of a geometrical progression in~\eqref{eq:charfunc_wrap_0} amounts to
\begin{equation}\label{eq:sum_geom_progr}
    \frac{1}{N}\sum_{n=0}^{N-1} e^{j2\pi \frac{p}{N} n} = \left\{
    \begin{array}{l l}
        1, & \quad \text{if } p\operatorname{mod}N=0, \\
        0, & \quad \text{else.}
    \end{array} \right.
\end{equation}
Finally, we obtain the characteristic function
\begin{equation}\label{eq:charfunc_wrap}
    \varphi_p' = \left\{
    \begin{array}{l l}
        e^{jp\mu} \frac{I_{p/N}(\kappa)}{I_0(\kappa)}, & \quad \text{if } p\operatorname{mod}N=0, \\
        0, & \quad \text{else.}
    \end{array} \right.
\end{equation}

Given the properties of the $N$-fold wrapped VM pdf, we choose the approximating MVM pdf
\begin{equation*}
    \frac{1}{2\pi I_0(\tilde\kappa) N} \sum_{n=0}^{N-1} e^{\tilde\kappa\cos(\theta - \tilde\mu - 2\pi n/N)},
\end{equation*}
i.e., the pdf has $N$ components with equal amplitudes, evenly spaced means $\tilde\mu + 2\pi n/N$, $n=0,\ldots,N-1$, and concentration parameters equal to $\tilde\kappa$. The characteristic function $\varphi_p''$, $p\in\mathbb{Z}$, is obtained as
\begin{align}
    \varphi_p'' &= \int_0^{2\pi} e^{jp\theta} \frac{1}{2\pi I_0(\tilde\kappa)N} \sum_{n=0}^{N-1} e^{\tilde\kappa\cos(\theta - \tilde\mu - 2\pi n/N)} \operatorname{d}\!\theta \nonumber \\
    &= \frac{e^{jp\tilde\mu}}{2\pi I_0(\tilde\kappa)} \frac{1}{N} \sum_{n=0}^{N-1} e^{j 2\pi \frac{p}{N}n} \int_0^{2\pi} e^{jp\theta} \, e^{\tilde\kappa\cos\theta} \operatorname{d}\!\theta \nonumber
\end{align}
Using~\eqref{eq:sum_geom_progr} again, we obtain
\begin{equation}\label{eq:charfunc_mixt}
    \varphi_p'' = \left\{
    \begin{array}{l l}
        e^{jp\tilde\mu} \frac{I_{p}(\tilde\kappa)}{I_0(\tilde\kappa)}, & \quad \text{if } p\operatorname{mod}N=0, \\
        0, & \quad \text{else.}
    \end{array} \right.
\end{equation}

We want to find $\tilde\mu$ and $\tilde\kappa$ that provide a good match between $\varphi_p'$ and $\varphi_p''$, for all $p\in\mathbb{Z}$.
Setting $\tilde\mu = \mu$, we obtain $\arg\{\varphi_p'\} = \arg\{\varphi_p''\}$, for all $p$. As to the magnitudes, we equate the first nonzero values of the characteristic functions, i.e., $\varphi_N'=\varphi_N''$, and obtain the transcendental equation in $\tilde\kappa$
\begin{equation}\label{eq:match_charfN}
    \frac{I_N(\tilde\kappa)}{I_0(\tilde\kappa)} = \frac{I_1(\kappa)}{I_0(\kappa)}
\end{equation}

To show that~\eqref{eq:match_charfN} yields a good approximation $|\varphi_p''| \simeq |\varphi_p'|$ for any $p$, we make use of the fact that the VM distribution characterized by $\mu$ and $\kappa$ can be well approximated by a wrapped normal distribution with mean direction $\mu$ and mean resultant length $\rho = A(\kappa) \triangleq I_1(\kappa)/I_0(\kappa)$~\cite{MardiaJupp2000}. While the approximation is tight for large $\kappa$, it is still satisfactory for intermediate values of $\kappa$. Therefore, the characteristic function $e^{j\mu p} \rho^{p^2}$ of the wrapped normal distribution approximates that one of a VM distribution. Based on~\eqref{eq:charfunc_VM}, we can thus write:
\begin{equation}\label{eq:approx_VM_wrapN}
    \frac{I_p(\kappa)}{I_0(\kappa)} \simeq \rho^{p^2},
\end{equation}
for all $p\in\mathbb{Z}$ and $\rho = A(\kappa)$. Next, we define $\tilde\rho$ by $\tilde\rho^{N^2} = I_N(\tilde\kappa)/I_0(\tilde\kappa)$. According to~\eqref{eq:match_charfN}, $\tilde\rho^{N^2} = \rho$. Thus, using~\eqref{eq:approx_VM_wrapN}, we obtain that, for all $p\in\mathbb{Z}$, $p\operatorname{mod}N=0$,
\begin{equation*}
    \frac{I_{p/N}(\kappa)}{I_0(\kappa)} \simeq \rho^{p^2/N^2}
    \simeq (\tilde\rho)^{p^2}
    \simeq \frac{I_{p}(\tilde\kappa)}{I_0(\tilde\kappa)}.
\end{equation*}
In conclusion, setting $\tilde\mu = \mu$ and solving~\eqref{eq:match_charfN} for $\tilde\kappa$, we find a good approximation $\varphi_p'' \simeq \varphi_p'$ for all $p$ and thus can write
\begin{equation}\label{eq:approx_pdfs_WMV_MVM}
    f_\text{VM}(N\theta;N\mu,\kappa) \simeq \frac{1}{N} \sum_{n=0}^{N-1} f_\text{VM}(\theta;\mu + 2\pi n/N,\tilde\kappa).
\end{equation}

An approximate solution to~\eqref{eq:match_charfN} can be found by using~\eqref{eq:approx_VM_wrapN} to arrive at $[A(\tilde\kappa)]^{N^2} = A(\kappa)$, where $A(\cdot)=I_1(\cdot)/I_0(\cdot)$.
Approximations of the function $A(\cdot)$  and its inverse are well studied, see~\cite[p. 40]{MardiaJupp2000} and~\cite[pp. 85--86]{MardiaJupp2000}.

\bibliographystyle{IEEEtran}
\bibliography{IEEEabrv,References}

\begin{IEEEbiographynophoto}{Mihai-Alin Badiu}
received the Dipl.-Ing., M.S. and Ph.D. degrees in electrical engineering from the Technical University of Cluj-Napoca, Romania, in 2008, 2010 and 2012, respectively. Since 2012, he has been with the Department of Electronic Systems, Aalborg University, Denmark, where he is currently holding a post-doc fellowship from the Danish Council for Independent Research. From 2008 to 2010 he was a research assistant at the Technical University of Cluj-Napoca. In 2011 he was a visiting PhD researcher at Aalborg University, Denmark. In 2016--2017 he was a visiting postdoctoral researcher at Aston University, Birmingham, United Kingdom. His research interests are in the fields of machine learning, wireless networks and signal processing.
\end{IEEEbiographynophoto}

\begin{IEEEbiographynophoto}{Thomas Lundgaard Hansen}
received the B.Sc. and M.Sc. (cum laude) in electrical engineering from Aalborg University, Denmark, in 2011 and 2014, respectively. Since 2014 he has been a Ph.D. fellow at Aalborg University. During 2013 and 2015 he was a visiting scholar at University of California, San Diego. He is the recipient of the best student paper award at the 2014 IEEE Sensor Array and Multichannel Signal Processing workshop and also received an award from IDA Efondet for his masters thesis. His research interests include signal
processing, machine learning and wireless communications.
\end{IEEEbiographynophoto}

\begin{IEEEbiographynophoto}{Bernard Henri Fleury}
(M'97--SM'99) received the Diplomas in electrical engineering and in mathematics in 1978 and 1990 respectively and the Ph.D. degree in electrical engineering in 1990 from the Swiss Federal Institute of Technology Zurich (ETHZ), Zurich, Switzerland.

Since 1997, he has been with the Department of Electronic Systems, Aalborg University, Aalborg, Denmark, as a Professor of Communication Theory. From 2000 till 2014 he was Head of Section, first of the Digital Signal Processing Section and later of the Navigation and Communications Section. From 2006 to 2009, he was partly affiliated as a Key Researcher with the Telecommunications Research Center Vienna (ftw.), Vienna, Austria. During 1978--1985 and 1992--1996, he was a Teaching Assistant and a Senior Research Associate, respectively, with the Communication Technology Laboratory, ETHZ. Between 1988 and 1992, he was a Research Assistant with the Statistical Seminar at ETHZ.

Prof. Fleury's research interests cover numerous aspects within communication theory, signal processing, and machine learning, mainly for wireless communication systems and networks. His current scientific activities include stochastic modeling and estimation of the radio channel, especially for large systems (operating in large bandwidths, equipped with large antenna arrays, etc.), especially when these systems operate in harsh conditions, e.g. in highly time-varying environments; iterative message-passing processing, with focus on the design of efficient feasible architectures for wireless receivers; localization techniques in wireless terrestrial systems; and radar signal processing. Prof. Fleury has authored and coauthored nearly 150 publications and is co-inventor of 6 filed or published patents in these areas. He has developed, with his staff, a high-resolution method for the estimation of radio channel parameters that has found a wide application and has inspired similar estimation techniques both in academia and in industry.
\end{IEEEbiographynophoto}

\end{document}